\documentclass[aaspp4]{aastex}
\usepackage{epsfig,rotating,amsmath,graphics}
\slugcomment{Accepted by \it The Astronomical Journal\rm}

\shortauthors{HIPASS HVCs}
\shorttitle{Putman et~al.}

\def\gtrapprox{\;\lower 0.5ex\hbox{$\buildrel >\over \sim\ $}}
\def\lessapprox{\;\lower 0.5ex\hbox{$\buildrel < \over \sim\ $}}          
\def\Msun  {${\rm M}_\odot$}
\def\deg   {$^\circ$}
\def\arcdeg {$^\circ$}

\def\HI    {H{\sc I}~}

\def\etal  {{et al.}}
\def\kms   {km s$^{-1}$}
\def\deg{\hbox{$^\circ$}}
\newfont{\tf}{cmr10 scaled 620}
\newfont{\ttf}{cmtt10 scaled 620}
\newcommand{\mtf}{\scriptstyle}

\renewcommand{\arraystretch}{0.75}

\begin{document}

\title{HIPASS High--Velocity Clouds: Properties of the Compact and Extended
Populations}

\author{
M.E. Putman,\altaffilmark{1,2,3}
V. de Heij,\altaffilmark{4}
L. Staveley--Smith,\altaffilmark{2}
R. Braun,\altaffilmark{5}
K.C. Freeman,\altaffilmark{1}
B.K. Gibson,\altaffilmark{6}
W.B. Burton,\altaffilmark{4}
D.G. Barnes,\altaffilmark{6}
G. D.  Banks,\altaffilmark{7}   
R. Bhathal,\altaffilmark{8} 
W. J. G. de Blok,\altaffilmark{2} 
P. J. Boyce,\altaffilmark{7} 
M. J. Disney,\altaffilmark{7} 
M. J. Drinkwater,\altaffilmark{9}
R. D. Ekers,\altaffilmark{2} 
P. A.  Henning,\altaffilmark{10} 
H. Jerjen,\altaffilmark{1}
V. A. Kilborn,\altaffilmark{11}
P. M. Knezek,\altaffilmark{12}
B. Koribalski,\altaffilmark{2} 
D. F. Malin,\altaffilmark{13}
M. Marquarding,\altaffilmark{2}
R. F. Minchin,\altaffilmark{7}  
J. R. Mould,\altaffilmark{1} 
T. Oosterloo,\altaffilmark{5} 
R. M. Price,\altaffilmark{10} 
S. D. Ryder,\altaffilmark{13} 
E. M. Sadler,\altaffilmark{14}  
I. Stewart,\altaffilmark{2,15} 
F. Stootman,\altaffilmark{8} 
R. L. Webster,\altaffilmark{9}
and A. E. Wright\altaffilmark{2}
}

\altaffiltext{1}{Research School of Astronomy and Astrophysics, 
ANU, Weston Creek P.O., Weston, ACT 2611, Australia.}

\altaffiltext{2}{Australia Telescope National Facility, CSIRO, P.O. Box 76,
Epping, NSW 2121, Australia}

\altaffiltext{3}{Currently a Hubble Fellow at: Center for Astrophysics and Space Astronomy, University of 
Colorado, Boulder, CO 80309-0389, USA. email: mputman@casa.colorado.edu}

\altaffiltext{4}{Sterrewacht Leiden, P.O. Box 9513, 2300 RA Leiden, The Netherlands}

\altaffiltext{5}{Netherlands Foundation for Research in Astronomy, P.O. Box 2, 7990 AA Dwingeloo, The Netherlands}

\altaffiltext{6}{Centre for Astrophysics \& Supercomputing, Swinburne University, Mail \#31, P.O. Box 218, Hawthorn, VIC, Australia 3122}

\altaffiltext{7}{University of Wales, Cardiff, Department of Physics 
\& Astronomy, P.O. Box 913, Cardiff CF2 3YB, U.K.}

\altaffiltext{8}{University of Western Sydney Macarthur, Department of 
Physics, P.O. Box 555, Campbelltown, NSW 2560, Australia}

\altaffiltext{9}{School of Physics, University of Melbourne, 
Victoria 3010, Australia}

\altaffiltext{10}{University of New Mexico, Department of Physics \& 
Astronomy, 800 Yale Blvd. NE, Albuquerque, NM 87131, USA}

\altaffiltext{11}{University of Manchester, Jodrell Bank Observatory, 
Lower Withington, Macclesfield, Chesire, England, SK11 6QS}

\altaffiltext{12}{Space Telescope Science Institute, 3700 San Martin
Drive, Baltimore, MD, 21218, USA}

\altaffiltext{13}{Anglo-Australian Observatory, P.O. Box 296, 
Epping, NSW 1710,  Australia}

\altaffiltext{14}{University of Sydney, Astrophysics Department, School 
of Physics, A28, Sydney, NSW 2006, Australia}

\altaffiltext{15}{Department of Physics \& Astronomy, University of
Leicester, Leicester LE1 7RH, U.K.}

\begin{abstract}
   A catalog of Southern anomalous--velocity \HI clouds at Decl.
   $<+2\deg$ is presented.  This catalog is based on data from 
   the \HI Parkes All--Sky Survey (HIPASS) reprocessed with the {\sc minmed5} procedure
   (Putman et al. 2002; Putman 2000), and searched with the high--velocity cloud
   finding algorithm described by de Heij et al. (2001). The improved
   sensitivity (5$\sigma$: $\Delta T_{\rm B}$ = 0.04 K), resolution
   (15\farcm5), and velocity range ($-500 < V_{\rm LSR} < +500$ \kms)
   of the HIPASS data,
   results in a substantial increase in the number of individual clouds
   (1956, as well as 41 galaxies) compared to what was known from earlier Southern data. The
   method of cataloging the anomalous--velocity objects is described,
   and a catalog of key cloud parameters, including velocity, angular
   size, peak column density, total flux, position angle, and degree of
   isolation, is presented.  The data are characterized into several
   classes of anomalous--velocity \HI emission.  Most
   high--velocity emission features, HVCs, have a filamentary
   morphology and are loosely organized into large complexes extending
   over tens of degrees. In addition, 179 compact and
   isolated anomalous--velocity objects, CHVCs, are identified based
   on their size and degree of isolation.  25\% of the CHVCs
   originally classified by Braun \& Burton (1999) are reclassified
   based on the HIPASS data.  
   The properties of all the high-velocity emission
   features and only the CHVCs are investigated, and distinct similarities and
   differences are found.  Both populations have typical \HI masses of 
   $\sim$ 4.5 D$_{\rm kpc}^2$ \Msun~and have  similar slopes for their column density 
and flux distributions. 
On the other hand, the CHVCs appear to be clustered and the population can be broken up into three 
   spatially distinct groups, while the entire population of clouds is more
   uniformly distributed with a significant percentage aligned
   with the the Magellanic Stream.  The median
velocities are $V_{\rm GSR} = -38$ \kms~for the CHVCs and $-30$ \kms~for
all of the anomalous--velocity clouds.  Based on the catalog sizes,
   high-velocity features cover 19\% of the southern sky, and CHVCs cover 1\%.

\end{abstract}


\section{Introduction}

Evidence for diffuse intergalactic material comes directly from
intra-cluster X-ray emission at high-redshift (Mulchaey 2000) and the Ly$\alpha$ 
absorber systems (Penton, Shull, \& Stocke 2000).   However, the amount of
intergalactic neutral hydrogen at $z=0$ remains highly uncertain,
and its presence has important
implications on the efficiency and timescales of galaxy formation 
and the epoch of reionization (Gnedin 2000; Rees 1986).  If 
primordial concentrations of neutral gas exist, they could
also house a large fraction of the missing mass in the universe
(Moore et al. 1999; Klypin et al. 1999).

Anomalous--velocity concentrations of atomic hydrogen surround our Galaxy,
and the formation and evolution of our Galaxy is evidently strongly linked
to an environment which includes these objects, as well as the Galaxy's extended massive halo and the galaxies of the Local
Group.  These atomic hydrogen concentrations
have velocities forbidden by simple models of Galactic rotation and
the debate continues over their distance and role in galaxy formation. Our ability to unravel
their origin is
complicated by the bulk classification of essentially all
anomalous--velocity \HI into a single type of object which does
not conform with Galactic rotation and does not contain stars.  In
reality there are several classes of this anomalous emission (e.g. see
 Wakker \& van Woerden 1997, Putman \& Gibson
1999), including high--velocity cloud complexes located within
some 10 kpc of the Galactic plane, tidal streamers associated with the
gravitational interaction of the Magellanic Clouds and the Milky Way,
and compact, isolated clouds which may be scattered throughout the
Local Group.  The origins and physical relevance of anomalous--velocity
\HI emission can be more readily determined once a clear classification into
sub--categories has been made.

The possibility that some of the anomalous--velocity clouds are
distributed throughout the Local Group has been considered earlier, by
(among others) Oort (1966, 1970, 1981),
Verschuur (1975), Eichler (1976), Einasto et
al.~(1976), Giovanelli (1981), Bajaja et al.~(1987), Wakker \& van Woerden (1997), Blitz et
al.~(1999), and Braun \& Burton (1999).  Blitz et
al.~(1999) showed that several general properties of the
anomalous--velocity \HI sky could be interpreted in terms of the
hierarchical structure paradigm for the formation and evolution of
galaxies.  In this context, the extended, filamentary \HI complexes,
high--velocity clouds (HVCs), would be relatively nearby objects
currently undergoing accretion onto the Galaxy or interacting with one
of the Magellanic Clouds, while the compact, isolated objects, (CHVCs,
following Braun \& Burton 1999: hereafter BB99) would be their
less evolved counterparts scattered throughout the Local Group environment,
typically at distances of several hundred kpc and with \HI masses of
about $10^{6-7}$ \Msun.

BB99 stressed that the population of anomalous--velocity \HI in the
Northern hemisphere can be split into distinct CHVC and HVC
components when observed with the spatial sampling and sensitivity of
the Leiden/Dwingeloo Survey of Hartmann \& Burton (1997).  The
Southern sky, however, has remained severely undersampled, leaving a
major gap in our knowledge of the distribution of the compact clouds.
Although HVC features are distributed over the entire sky
(see Wakker \& van Woerden 1997), most studies have focused on
the Northern complexes due to limited Southern neutral hydrogen data.
This paper characterizes the high--velocity \HI sky south of
Decl. $+2$\deg~ and within the {\sc lsr} (Local Standard of Rest) velocity range $-500$
\kms~to +500 \kms, using data from the HI Parkes All-Sky Survey
(Barnes et al. 2001), reduced with an algorithm which recovers
extended emission (Putman et al. 2002; Putman 2000). The data
allow the highest spatial resolution ($15\farcm5$) large--scale study
of the anomalous--velocity HVC and CHVC phenomena to date. Since these
clouds show an increasingly complex structure at higher resolution
(e.g. Wakker \& Schwarz 1991, Braun \& Burton 2000),
HIPASS has great potential for providing new insights into
these objects.  The positional accuracy of HIPASS will also encourage
further distance and metallicity determination work to be carried out
in the Southern hemisphere.

In this paper, we describe a catalog of the anomalous--velocity clouds
detected by HIPASS which includes the distinction between
compact and extended objects.  The method of cataloging the clouds is described, and the
spatial and kinematic distributions of high--velocity clouds and compact
high--velocity clouds are discussed separately.  The flux, column density,
size, and position angle
properties of the populations are also thoroughly investigated.

\section{Data and catalog}

\subsection{Observations}

The neutral hydrogen data are from the \HI Parkes All--Sky Survey
(HIPASS) reduced with the {\sc minmed5}
method (Putman et al. 2002; Putman 2000).  HIPASS is a survey for \HI in the
Southern sky, extending from the South celestial pole to Decl.
$+2\deg$, over velocities from $-1280$ to $+12700$ km s$^{-1}$ (see Barnes 
et al. 2001 for a full description).  The
survey utilized the 64--m Parkes radio telescope, with a focal--plane
array of 13 beams arranged in a hexagonal grid, to scan the sky in
$8^\circ$ zones of Decl. with Nyquist sampling.  The
spectrometer has 1024 channels for each polarisation and beam, with a
velocity spacing of 13.2 km s$^{-1}$ between channels and a spectral
resolution, after Hanning smoothing, of 26.4 km s$^{-1}$.  Each 8\deg $\times$
1.7\deg~ scan
was re--observed five times (460 seconds/beam total integration time) 
to reach the full survey sensitivity.  This
repetitive procedure provides source confirmation, mitigates diurnal
influences and aids interference excision.

HIPASS was designed to detect discrete \HI sources.  In fact, the
standard HIPASS reduction method described by Barnes et al.~(2001)
filters out emission that extends over more than 2\deg~ of Decl.
This is because a running median filter of this extent is used to provide a local
bandpass calibration of each spectrum.  Many of the
anomalous--velocity \HI objects are extended on larger angular scales,
so an alternate method of bandpass calibration was developed (known as 
the {\sc minmed5} method) which recovers emission which extends 
up to 7\deg~ in Decl.  {\sc minmed5} calculates the
bandpass correction for each beam, polarization and velocity by
first dividing the 8\deg~ scan into five sections, then finding the
median emission in each section, and subsequently using the minimum of
the five values. The {\sc minmed5} procedure greatly increases the
sensitivity of the data to large--scale structure without substantial
loss of flux density, except when the emission fills the entire
8\deg~ scan.  This circumstance occurs for \HI emission from
the Galactic Plane and from the Magellanic Clouds.  \HI emission in 
the {\sc lsr} velocity range $-700$ to
$+1000$ \kms~was reduced in this manner.  Further details of the {\sc
minmed5} procedure are given by Putman et al.~(2002) and Putman (2000).

The calibrated scans were gridded with the median method
described by Barnes et al.~(2001), without the weighting
which overcorrects the fluxes for extended sources. The median method has the
desirable effect of being robust to intermittent interference, but
causes both the effective beam area as well as
the peak gridded response to vary systematically with source
size. Completely unresolved sources have a response reduced by a factor
of 1.28, while sources larger than about 40~arcmin have the nominal
response (as tabulated in Barnes et al.). We have taken account of this
effect in our analysis.  
For unresolved sources the {\sc rms}, determined from the data cubes
   using an iterative 3-$\sigma$ clip, is $13\pm1$ mJy beam$^{-1}$.
   Although nominally the same as the Barnes et al. HIPASS sensitivity,
   it is slightly worse than expected, considering the greater velocity 
   smoothing (26 km s$^{-1}$ cf. 18 km s$^{-1}$) in the current data.
   The difference appears due to the closer proximity in velocity to bright
Galactic features. For extended sources, the RMS noise is
   10 mJy beam$^{-1}$ (beam area 243 arcmin$^2$), corresponding to
   a brightness temperature sensitivity of 8 mK.
The spatial size of the gridded cubes is 24\deg~ $\times$ 24\deg~ with
a few degrees of overlap between each cube.

\subsection{Cloud search algorithm and selection criteria}

The cataloged clouds were initially defined using data cubes which were
smoothed spectrally by a Gaussian with a FWHM of 31 \kms, and
spatially with a Gaussian with a FWHM of 19\arcmin.  The cloud
parameters listed in the catalog (and in Table 1) were subsequently
measured from the unsmoothed cubes (15\farcm5 spatial resolution and 26
\kms~velocity resolution). The entire velocity range $-700 < V_{\rm
LSR} < +1000$ \kms~of the cubes was searched, excluding velocities $|V_{\rm
LSR}|<90$ \kms\ where confusion with emission from the conventional
Milky Way gaseous disk is difficult to avoid.  The catalog only includes objects
between $\pm 500$ \kms, since no high-velocity cloud was found beyond this velocity range.  The galaxies found beyond that range will be included
in the HIPASS Bright Galaxy Catalog (Koribalski et al. 2002).
All of the detections within $\pm 500$ \kms\
which correspond with known galaxies recorded in the LEDA and NED
databases are included in the table and appropriately labeled, but are
excised from the figures.  A total of 1997 anomalous--velocity \HI
objects are represented in the catalog, of which 41 are galaxies (one of
which was previously unknown), 1618 are HVCs, 179 are CHVCs and 159 are 
designated :HVC (see Section 2.4).
The median properties of the four categories of cataloged objects
are listed in Table 2.

A full description of the cloud search algorithm is given by de~Heij et
al. (2001).  The steps used to catalog the HIPASS clouds are
briefly summarised as follows:

\begin{enumerate}
\item All pixels with $T_{\rm B} > 6$ mK in the smoothed datacubes were
examined, and each was assigned to a local maximum with which it is
connected (spatially and in velocity; there are 26 neighboring pixels).
\item Only those local maxima with $T_{\rm B}$(max)$ > 12$ mK were retained.
\item Adjacent local maxima were merged into a single cloud if the
brightest enclosing contour for the particular maximum had $T_{\rm B} >
80$ mK (i.e. if there exists a bright connection between the maxima) or
if $T_{\rm B} > 0.4 T_{\rm B}$(max) (i.e. if the contrast between the
local maxima is small).
\item Finally, the merged maxima were only labelled as clouds if
$T_{\rm B}$(max) $> 5\sigma$, where
$\sigma$ is the locally defined noise level at the velocity of each
cloud candidate.  $\sigma$ was determined from the 
clipped {\sc rms} of pixel values within a $4\deg~ \times 4\deg~$
box. Pixel values were clipped if they deviated by more than 3 times
the absolute deviation from the median pixel value.  The minimum noise value is
8 mK.
\end{enumerate}

After following the above process to determine which pixels are part of
a cloud, the unsmoothed data cubes were used to create
integrated--intensity images which include the entire range of
velocities of the pixels assigned to the cloud.
Fig.~\ref{fig:examples} shows two representative integrated--intensity
maps of clouds cataloged in the above manner. A rectangular box
defines the maximum spatial extent of the pixels assigned to each
cataloged object in the figure. Due to the loss of the velocity
information, the projected distribution shows the objects to
overlap, but each pixel is assigned to only one
object. Each integrated--intensity map and the spectrum in the
direction of object's peak brightness was also inspected by eye. A small number
of obvious artifacts (e.g. due to failed bandpass calibration) were
removed manually.

Finally, because the velocity cut of $|V_{\rm LSR}|<90$ \kms~does not
exclude all emission directly associated with the Milky Way towards
the Galactic Plane (see Hartmann \& Burton 1997), an additional selection criterion was applied to the catalog
candidates based on the deviation velocity,~$V_{\rm dev}$.  As defined
by Wakker (1991), the deviation velocity of a cloud is the smallest
difference between the velocity of the cloud and any allowed Galactic
velocity in the same direction.  The kinematic and spatial properties
of the conventional Galactic HI were modelled in order to determine the
deviation velocity from the Local Standard of Rest velocity.  The model
(see de Heij et al. 2001) consists of a thin disk which has constant properties (volume density,
scale--height, and temperature) within the solar radius, but flares and
warps further outward.  The gas exhibits circular rotation with a flat
rotation curve of 220 \kms.  The velocity is considered an acceptable
Galactic velocity as long as the intensity of the synthetic \HI spectra
at that position exceeds 0.5 K.  We required that $|V_{\rm
dev}|$ be greater than 60 \kms~for the feature to be included in the catalog.
Some 209 candidates were excluded from the catalog based upon this
criteria, or approximately 10\%.  The excluded objects probably include
some whose appearance is due to small--scale structure in the gaseous
disk of the Milky Way, but may also include bonafide HVCs or CHVCs, or
external galaxies.
Several HVC populations included in the Wakker \& van Woerden (1991;
hereafter WvW91) catalog have
$|V_{\rm dev}| <$ 60 \kms, and are thus excluded from this catalog (e.g.
Populations Co-rotating, OA, GCP).  Those objects which have emission at
high velocities, $|V_{\rm dev}| >$ 60 \kms, but which clearly link to gas
with $|V_{\rm dev}| <$ 60 \kms~are catalogued separately and referred to as
XHVCs (see Table 3).
There are 186 XHVCs and 4 galaxies which merge
with low velocity Galactic emission and are also included in the XHVC catalog
(NGC253, NGC292 (the Small Magellanic Cloud), NGC301, and IC5152).

\subsection{Catalog}

The results of the {\sc minmed5} reduced HIPASS data and the
anomalous--velocity cloud finder described above are given in the catalog
which is begun in Table 1.
The complete catalog can be found in the electronic version of the Journal
and at {\it http://www.atnf.csiro.au/research/multibeam}.  
As discussed in section 2.1, sources less than 40$^\prime$ in diameter have had their 
brightness corrected for based on the object's angular size, using the values
listed in table~3 of Barnes et al.
The catalog contains the following entries:
\begin{description}
\item Column~1:  HIPASS HVC running number.
\item Column~2: Designation, consisting of a prefix followed by the
Galactic longitude, Galactic latitude, and Local Standard of Rest
velocity.  The prefix is HVC for filamentary features and their
sub--structures which are organized into large high--velocity cloud
complexes; CHVC for those satisfying the criteria of compact, isolated,
high--velocity clouds (as discussed in Section 2.4); :HVC for clouds which could
not be unambiguously classified as a CHVC; and GLXY for galaxies
cataloged in either the LEDA or the NED databases or visible in Digital
Sky Survey (DSS) data. The DSS data was retrieved from the Space
Telescope Science Institute for a 30~arcmin diameter field centered on
each catalog entry with a {\sc gsr} velocity in excess of 100~\kms. Integer
rounded positions are used for larger sources, while an extra decimal
is used for the smaller ones. The Galactic longitude and latitude were
calculated from the intensity--weighted average of the RA
and Decl. (columns 3 and 4).
\item Columns 3 and~4: Intensity--weighted average (J2000) 
RA and Decl. of all pixels that are part of the object.
\item Column~5:  Intensity--weighted average Local Standard of Rest
velocity of all pixels in the object.
\item Columns 6 and 7: Average velocities, expressed in the Galactic
Standard of Rest and in the Local Group Standard of Rest frames.  The
Galactic Standard of Rest reference system is defined by $V_{\rm GSR} =
220\cos(b)\sin(l) + V_{\rm LSR}$; the Local Group Standard of Rest
system, by $V_{\rm LGSR} = V_{\rm GSR} - 62\cos(l)\cos(b) +
40\sin(l)\cos(b) - 35\sin(b)$, following BB99.
\item Column~8:  Velocity FWHM from the spectrum which passes through the
peak T$_{B}$ pixel of the object.  Those values below 26 \kms\ (the velocity
resolution) should be thought of as having 26 \kms\ as the upper limit.
\item Columns 9, 10 and~11: Angular semi--major axis, semi--minor
axis, and position angle of the Full Width, Half Power (FWHP) ellipse
in degrees.  Using the velocity--integrated column density
image of the object, an ellipse was fit to the contour with half the value of the
maximum column density (column~14).  Thus the major and
minor axis are not defined by the original pixels assigned
to the object, as the other parameters are, but from
the column density image.  In almost all cases, the area from the major and minor
axes is smaller than the area from the total number of pixels assigned to
the cloud, but in a few cases it is larger due to the column density image
including overlapping features.  These size values were
predominantly used to determine the designation.  The position angle is
positive in the direction of increasing Galactic longitude, with a
value of zero when the object is aligned pointing toward the Galactic
North pole.
\item Column~12:  The angular size based on the total number pixels assigned
to an object times the area of a $4^{\prime} \times 4^{\prime}$ pixel. 
\item Column~13: Peak brightness temperature in K.
\item Column~14: Maximum column density in units of $10^{20}$ cm$^{-2}$.
\item Column~15: Total flux, in units of Jy\,\kms.  
\item Column~16: Name of any previously cataloged galaxy in the LEDA or
NED databases. Newly discovered galaxies (based on the fact that
they are not in LEDA or NED) which have an optical counterpart in the DSS data have been
given a HIPASS J2000 name designation. If the object has a clearly
discernable optical counterpart it is followed by ``(A)'', if it does not,
and has a GSR velocity exceeding 300~\kms, it is followed with ``(B)''.

\end{description}

Among the objects cataloged, there will be a substantial number
which are not new discoveries.  We have not made cross--references to
earlier detections, except for those objects identified with
nearby galaxies named in the LEDA or NED compilations.  Firmly establishing cross--references to earlier work
would be a difficult matter, as in almost all cases, the angular resolution
and sensitivity of the HIPASS data are superior to earlier data.  Earlier
work established the general outlines of the major complexes, including the
Magellanic Stream.  Many of the HIPASS cataloged entries fall within these general
outlines, but are seen as separately indentifiable sub--structures at the
HIPASS resolution. 
Examples of just a few of the
cataloged objects which have been subject to substantial earlier work are given:

\noindent $\bullet$ HIPASS HVC \#91,
HVC\,$008.2-04.6-214$, and HIPASS HVC \#93, HVC\,$008.4-04.4-213$, were
discovered by Shane in Dwingeloo observations and reported by Oort (1968)
and Hulsbosch (1968) before being subject to a more extensive analysis,
based on new data, by Saraber \& Shane (1974; see also Mirabel \& Franco
1976).  Shane's object is identified as two objects in the
higher--resolution HIPASS material.  It is listed by WvW91 as their \#307.

\noindent $\bullet$ HIPASS HVC \#605,
CHVC\,$217.7+00.1+278$, has been discussed as Dw\,$217.8+0.0$ by Henning
(2000) and Rivers (2000) as a nearby galaxy hidden behind the Zone of
Avoidance and discovered in the Dwingeloo Obscured Galaxies Survey (see
Henning et al. 1998).  It is also listed by Henning et al.
(2001) as HIZSS \#003, at $l=217\fdg7,\,b=+0\fdg08$.  As no optical identification has yet been made, we retain
the object as a CHVC, though it could be one of many nearby galaxies masquerading as a CHVC.

\noindent $\bullet$  HIPASS HVC \#1914,
HVC\,$346.5+35.6-107$, was the subject of Westerbork synthesis imaging
carried out by Stoppelenburg, Schwarz, \& van Woerden (1998).  Stoppelenburg
et al. cross--reference object \#132 in WvW91's
tabulation, although that object is located some $2^\circ$ distant from the
HIPASS position, illustrating the difficulties with detailed
cross--referencing between catalogs with such different observational
parameters.  Besides the resolution difference for the southern part of the WvW91
catalog (35\arcmin\ beam on a 2\deg~ $\times$ 2\deg~ grid), they also 
merged clouds by examining the individual components by eye for a 
spatial connection at similar velocities.


\subsection{Definition of a compact, isolated high--velocity cloud}

The early surveys for high--velocity clouds were based on rather coarse sampling of the
sky and only modest sensitivity and this made it difficult to assess the cloud's
morphology, or to make more than
the simplest organization of objects into ``complexes''.  With the
advent of more finely sampled, sensitive surveys it has become possible
to utilize source morphology in the process of classification.  BB99
investigated the 
Leiden/Dwingeloo Survey (LDS) of Hartmann \& Burton (1997) and
found a dichotomy in the
types of \HI\ emission features seen at anomalous velocity. The large
majority of high--velocity emission originates in relatively diffuse
filaments which are organized into complexes extending over large
regions of the sky.  There is sub--structure in
these filaments, but it is generally consistent with only modest changes in the
viewing geometry and the line--of--sight overlap of a population of
intrinsically elongated structures extending over many degrees on the
sky.
In addition, they recognized a second type of high-velocity cloud,
namely compact, isolated objects with a high
column--density contrast.  The \HI\ signatures of these
objects are indistinguishable from those of many
dwarf galaxies.  
As stated in the introduction, BB99 refer to this subclass of 
compact, isolated clouds as CHVCs and argue that the kinematic and
spatial properties of the CHVCs are suggestive of Local Group distances
and a net infall velocity of about 100 \kms~towards the barycenter
of the Local Group.
 Since the sample of BB99 was drawn
from the Leiden/Dwingeloo Survey (by eye), their study did not include clouds south
of Decl.$-30$\deg, nor more detail than afforded by the
35\arcmin\ resolution of the data.

The spatial resolution and sky coverage of the {\sc minmed5} processed
HIPASS make it an ideal dataset for the study of CHVCs in the Southern
sky. The combination of high--resolution, Nyquist--sampled imaging and
high sensitivity reveal distinct morphological features associated
with the different categories of high--velocity \HI. 
The Magellanic
Stream, for example, is characterized by extremely long filaments
(typically $>10^\circ$ in extent) with a preferred alignment (parallel
to the great circle formed by RA $=23^{\rm h}$ and 11$^{\rm h}$),
while the extended accumulations at positive {\sc lsr} velocities in the
third longitude quadrant first observed by Wannier \& Wrixon
(1972) and Wannier, Wrixon, \& Wilson (1972) have a
completely different ``texture'', consisting of shorter filaments
(typically $5^\circ$ to $10^\circ$) with a wide range of
orientations. The CHVCs stand out from this filamentary
background, more so than with the coarser resolution of the
LDS. It seems likely that the distinctive resolved morphologies of
different anomalous--velocity objects might be used to better classify
those objects which share a common physical origin.
Here only the basic distinction between the filamentary HVC complexes 
and the CHVCs is examined.

The selection criteria for a CHVC was applied to a
$10^\circ \times 10^\circ$
velocity--integrated intensity image centered on each object, including
all pixels with a brightness $>$ 6~mK.
  The overriding consideration for the CHVC classification was the
brightness distribution at a level corresponding to about 25\% of the
peak column density of each object.  It was required that this contour
(1) be closed with an angular size less than 2$^\circ$; (2) not be
elongated in the direction of any lower level extended emission; and (3) be
more than $3\sigma$ above the local noise floor, so that the assessment
of elongation is meaningful.  In this way we hoped to avoid both the
low--contrast local maxima found along filamentary features as well as
local maxima formed by the line--of--sight overlap of several
filaments.  Sources clearly satisfying all three requirements were
designated CHVC in the catalog.  Those sources which
satisfied the size and signal--to--noise requirements, but could not be
unambiguously classified with regard to their elongation were
designated :HVC, while all remaining sources were given the HVC
designation.   Although the first and last of these conditions are
straightforward to apply, the second is difficult to incorporate into a
digital algorithm and was therefore applied interactively.  Since some
subjectivity is involved in this assessment, two of the authors (VdH and
RB) each
independently carried out a complete classification of all
sources.  Identical classification was given to 98.9\% of the 
sample, suggesting that the criterion
can be applied repeatably.   A consensus on classification was
reached on the remaining 25 sources after re-examination.  Fig. \ref{fig:grouping}
shows an example of typical objects classified as CHVCs, :HVCs and HVCs
respectively.

\subsection{Completeness of the Catalog}

Each cloud is automatically confirmed in the HIPASS data because of
the repetitive nature of the survey and the overlap between adjacent
scans (see Barnes \etal\ 2001).  Therefore re-observation and confirmation
of each object, such as done by BB99, is not required.  
A recent analysis of the Arecibo Dual Beam Survey 
(Schneider \& Rosenberg 2000) has shown that full survey completeness 
for compact sources is only reached at significance levels exceeding about 8$\sigma$. 
In the case of the CHVCs, we have required, as noted above,
that the contour at 25\% of peak brightness must be above 3$\sigma$ in
the velocity integrated image, so that the peak column density must
exceed 12$\sigma$. Since the typical CHVC FWHM linewidths are only
marginally resolved at 26 \kms, this corresponds to a limiting peak
column density of about 6$\times10^{18}$\,cm$^{-2}$ for CHVC detection.
The corresponding flux density limits for unresolved sources are
$\sim 2$ Jy~\kms\ for HVCs and $\sim 4$ Jy~\kms\ for CHVCs.
For mainly unresolved galaxies of low velocity width in the south celestial
cap region, Kilborn, Webster \& Staveley-Smith (1999) quote a similar HIPASS
completeness limit of 4 Jy~\kms.


The nature of the cloud cataloging algorithm, however, leads to
surprisingly poorer completeness and reliability levels than expected from the above
discussion.  Although many low flux-density galaxies are represented
in the catalog, comparison with the literature shows a systematic
tendency to underestimate the total flux density for objects with 
measured values as high as
$\sim 25$ Jy~\kms.  Completeness and reliability for extended clouds
is also poorer, as these objects cover many pixels. Typically,
truncation of the low column-density wings occurs, leading to an
underestimate of total flux density. Our present understanding of
the completeness limits are that the catalog is complete above $\sim 25$ Jy~\kms\
in flux density and $\sim 10^{19}$~cm$^{-2}$ in column density.

The above discussion has assumed a uniform noise level and a
sufficiently low source density that confusion does not become a
problem for source recognition and detection. While this is appropriate
for a large fraction of the position--velocity volume of the data,
there are also large regions which are adversely effected both by an
increased noise level and source confusion. The most prominent of these
is the gap between $V_{\rm dev} = -60$ \kms~and $+60$ \kms~({\sc lsr}),
which we leave entirely out of consideration due to confusion with
Galactic emission.  In addition, the data volumes occupied by the
Large and Small Magellanic Clouds, the Magellanic Stream, and
the major HVC complexes, give rise to an effective
increase in the {\sc rms} fluctuation level and a reduced ability to
detect potentially unrelated anomalous--velocity objects that overlap
in the measurement volume.   Therefore, we note that the catalog
may not be complete to the stated levels in these regions.

This catalog will be combined with a new catalog of the Northern sky to
obtain a complete all-sky CHVC catalog (de Heij et al. 2001).  However, the issue of
the superior spatial resolution, but
inferior velocity resolution, of the HIPASS material compared to that of
earlier surveys does not
make the combination immediately straight-forward.  To obtain a
uniform catalog with similar sensitivities and velocity information, it may be required to spatially smooth the HIPASS data and kinematically smooth the Northern
sky data.

There are 30 of the 65 CHVCs in the BB99 catalog which are found between Decl. -30\deg~ to +02\deg~ and are
therefore also cataloged here.  In general,
the position, velocity, FWHM, peak brightness temperature, and column
density properties of the clouds in both catalogs agree.  The largest
difference is the degree of isolation, or the CHVC/:HVC/HVC designation,
with 50\% of the BB99 CHVCs in the region of overlap reclassified as HVCs
or :HVCs.  The central velocities of the two catalogs agree within 10
\kms, and the FWHMs closely match when the LDS is
smoothed to HIPASS velocity resolution.
The average difference in the size of the major axis is
20\arcmin, however the size is based on the 50\% T$_{\rm max}$
contour in BB99 (all the resolution allowed for) and the 25\% T$_{\rm max}$
contour here.
T$_{\rm max}$ agrees within 32\% and N$_{\rm HI}$ within 45\%, with the
HIPASS value generally higher in both cases.
These results
make sense when one considers the different resolutions of the
surveys.  The improved spatial resolution of HIPASS changes the size of the
cloud, and its decrease in velocity resolution and increased
sensitivity may both be partially responsible for the change in isolation criteria.
The variation in column density for HVC sightlines at different
resolutions is expected, and this effect has been fully discussed by
Wakker, Oosterloo, \& Putman (2001) and Wakker \etal\ (2001).  
The positional accuracy was tested by comparing the published positions of
the galaxies in the catalog to the HIPASS values.  The positions generally
agree within 2\arcmin\ and almost always within 10\arcmin.

\section{Results} 
\subsection{General distribution properties}

Column density images of the high--velocity \HI sky are shown in polar--cap
projection, for three representative ranges of anomalous--velocities,
in Figs.~\ref{fig:hi_pos1}, \ref{fig:hi_pos2}, and \ref{fig:hi_neg}.
The Magellanic Clouds, portions of the Magellanic Stream, and the
Milky Way gaseous disk dominate the perceived fluxes, and are labelled
in the figures.  The emission
from the Galactic Plane is not cataloged, but the LMC and several low-velocity 
galaxies are cataloged and identified. The SMC merges with low velocity
Galactic emission (gas with $V_{\rm dev} < 60$ \kms) and is only included
in the catalog of XHVCs (Table 3).
Several properties of the high velocity gas are evident in Figs. 3$-$5,
including:  the broad range of spatial structures of the high-velocity 
features, the difficulty in defining a HVC based on the LSR definition
near the Galactic Plane, and the clear dichotomy in the location of 
the positive and negative LSR velocity features (negative predominantly at $\ell <
180$\deg~ and positive at $\ell > 180$\deg).

The spatial distribution of all of the cataloged HIPASS
anomalous--velocity objects is shown, in the same polar--cap
projection, in Fig.~\ref{fig:ra_dec} with the objects GSR velocity
represented.  No distinction is made in this
plot between objects identified in the catalog as HVCs, :HVCs, or
CHVCs, and the GLXYs are not included.  The figure shows local minima in the density of objects along
the Galactic disk and toward the Magellanic Clouds, where the survey
completeness is expected to be degraded (as discussed in Section 2.5).  
The band of objects extending from near
(RA, Decl.)$=(0^{\rm h},0^\circ$) to (4$^{\rm h},-50^\circ$)
contains the Magellanic Stream and the highest negative GSR velocities. The
large positive--velocity accumulation (designated WA, WB, WC, and WD by
WvW91) first noted by Wannier \& Wrixon
(1972) and Wannier et al.~(1972) is concentrated
between (RA, Decl.)$=(9^{\rm h} \rightarrow 12^{\rm h}, -10^\circ
\rightarrow -50^\circ$).  This region also contains material which
is most likely related to the Magellanic Leading Arm (Putman \etal\ 1998),
as shown in Fig. \ref{fig:hi_pos2}.
Other known complexes evident in this figure
include Complex L, centered near (RA, Decl.)$=(15^{\rm h},
-30^\circ)$; parts of the GCN and GCP Complexes, near
(RA, Decl.)$=(21^{\rm h}, -10^\circ)$; and the tail of Complex
C, near (RA, Decl.)$=(17^{\rm h}, -10^\circ)$. The GCN and GCP
Complexes appear to be part of an extension of clouds which
extend through the South Galactic Pole, perpendicular to the
Magellanic Stream.  A striking
feature of the distribution is the paucity of anomalous--velocity \HI in
the region bounded approximately by (RA, Decl.)$=(2^{\rm h}
\rightarrow 5^{\rm h}, -10^\circ \rightarrow -30^\circ)$.

The perceived structure of many of the concentrations of anomalous--velocity
HI emission is dependent on the projection used.  The distribution of
the 1956 cataloged clouds in Galactic coordinates is shown
in Fig.~\ref{fig:gal}, with the spatial limits of the catalog
represented by the dotted line.  The distribution of clouds is
similar to that presented by WvW91 (and also to the distribution of
high-velocity components presented by Morras \etal\ (2000)),
but there is a factor of $\sim 4$ increase in the number of
individual clouds south of Decl. 2\deg~ compared to the 
WvW91 catalog.  This increase in the number of
individual clouds is due to both the improvement in spatial resolution
and the catalog method.

The kinematic distribution of all of the cataloged anomalous--velocity
objects (i.e. HVCs, :HVCs, and CHVCs) is plotted against Galactic
longitude and latitude, for the Local, Galactic, and Local Group
reference frames in Figs. \ref{fig:glon_vlsr} and \ref{fig:vlsr_glat},
respectively. As noted previously, the majority of the Southern anomalous--velocity objects
at $l > 180\deg~$ are at positive {\sc lsr} velocities, and though the
number of objects at $l < 180\deg~$ is small, they are predominantly
at negative {\sc lsr} velocities. 
  This follows the trend of the rotating Galactic
\HI\ (Burton 1988), but of course the magnitude is much greater.  
 The largest range of velocities in every reference frame is at $l = 0$\deg.  In fact, some of these clouds become more
positive or negative in the GSR frame.  The distribution of the clouds in the
Local Group reference frame (Fig.~\ref{fig:glon_vlsr}c) is not greatly
different from the Galactic reference frame; the clouds are slightly
more spread out and the broad distribution at $l = 0$\deg~
remains.  The number of clouds at each Galactic longitude is
presented in Fig.~\ref{fig:glon_vlsr}d, but note it is not complete
at all longitudes, as shown in Fig. \ref{fig:gal}.  The overabundance
of clouds around $l \approx$ 270\deg~ is primarily due to the Magellanic
Stream which is a prominent feature in Figs. \ref{fig:glon_vlsr} and \ref{fig:vlsr_glat}.  In Fig. \ref{fig:vlsr_glat} the Stream provides a large component of the negative--latitude
gas, particularly the concentration near ($V_{\rm GSR},b)=(-200$
\kms,$-70$\deg).  The Magellanic Leading Arm (Putman et al. 1998) is represented by
the clouds near the Galactic Plane which remain at high velocities.


Histograms of the various velocity distributions are
shown in Fig. \ref{fig:vlsr_histo}. The formal standard deviations of the
distributions are 115~\kms~in both the {\sc gsr} and the {\sc lgsr}
systems, compared to 185~\kms~in the {\sc lsr} frame. 
Blitz et al. (1999) found a similar decrease between the
{\sc lsr} and {\sc gsr} dispersions using the all-sky WvW91 catalog and excluding the Magellanic Stream.
Because of the large gap in survey coverage at low {\sc lsr} velocities,
the {\sc lsr} value is an upper limit.   The median {\sc lsr} velocity for the HIPASS catalog is 114 \kms, while the median in the {\sc gsr}
distribution is at approximately $-30$~\kms, and $-63$~\kms~ in the {\sc lgsr} system.  Of course, the high velocity
objects of the Northern hemisphere -- where negative velocities dominate --
should also be included in this analysis before conclusions are drawn
on the clouds' overall distribution.
The all-sky WvW91 catalog has median velocities of -90 \kms~({\sc lsr}),
-38 \kms~({\sc gsr}), and -58 \kms~({\sc lgsr}).  Mean velocities are -33 \kms~({\sc lsr}), 
-48 \kms~({\sc gsr}), and -46 \kms~({\sc lgsr}) for the all-sky WvW91 catalog,
and 44 \kms~({\sc lsr}), -33 \kms~({\sc gsr}), and -54 \kms~({\sc lgsr})
with the Southern HIPASS catalog.

The distributions of several properties of the cataloged objects are
shown in Figs.~\ref{fig:flux_histo} to \ref{fig:coverage},
and the median properties are tabulated in Table 2.  
The distribution functions of \HI flux, peak
column density, and angular size can all be described by power laws
over a limited range of values.  Each of these distributions shows a
turn-over at low values. For \HI flux, the flattening
of the slope at $\sim 20$ Jy~\kms~evident in Fig. \ref{fig:flux_histo} corresponds
to an \HI mass, M$_{\rm HI} \approx 4.5$ D$_{\rm kpc}^2$
\Msun. This value is consistent with the likely completeness limit noted earlier.
Fig. \ref{fig:nhi_histo} shows that the peak
column density flattens from a power law below $N_{\rm HI} \sim 2\times
10^{19}$ cm$^{-2}$. This is slightly higher than the
completeness limit noted above of $\sim 10^{19}$
cm$^{-2}$. Some of the flattening may be real, but a better understanding
of the selection limits is needed.  
The angular size distribution based on the total number of pixels assigned to
each cloud is shown in Fig. \ref{fig:size_histo}.  The peak is
between 0.2 $-$ 0.7 deg$^{2}$, which is significantly 
resolved with the $\sim0.07$ deg$^{2}$ survey resolution.
The turnovers seen in all the distributions in the log-log plots of
Figs~\ref{fig:flux_histo}-\ref{fig:coverage} appear consistent with the presently 
understood selection effects inherent in the catalog, making it possible that
there exists an underlying power law in the distribution of total flux density,
column density and angular size to lower limits than those probed in the present
catalog.


The differential \HI flux distribution, $f(F_{\rm HI}) \propto F_{\rm
  HI}^{-2.1}$ (Fig.~\ref{fig:flux_histo}) has a slope (above 30 Jy
\kms) slightly more negative than the limiting value of $-2$, below which the
total flux becomes dominated by the contribution from the low flux
end. WvW91 obtained a shallower power--law
dependence, $f(F_{\rm HI}) \propto F_{\rm HI}^{-1.5}$, in their
discussion of earlier survey data with coarser resolution. The
principal reason for this difference is likely to be the qualitative
organization of individual objects into larger clouds and complexes,
which will populate the high flux tail of the distribution at the
expense of the low flux peak.  If the distance to individual objects
were known it would be possible to construct the \HI mass distribution
function from our data. However, it is possible that the HVC distances
span more than two orders of magnitude, so we can not easily constrain
the intrinsic mass function shape at this time.

The peak column--density distribution function, $f(N_{\rm HI}) \propto
N_{\rm HI}^{-2.9}$, shown in Fig.~\ref{fig:nhi_histo}, implies that high
observed column densities are extremely rare. 
The distribution turns over near $N_{\rm HI} \sim 8\times 10^{18}$ cm$^{-2}$,
and the median peak column density (Table 2) is $9\times 10^{18}$
cm$^{-2}$.
These values may be dependent on the resolution and completeness limits.
The median FWHM linewidth is 36~\kms~with a dispersion of only 12~\kms,
making the distribution in linewidth effectively single-valued. Since
the median is only marginally in excess of the relatively coarse velocity
resolution, the measured linewidth must be regarded as an upper limit.

Figure~\ref{fig:size_histo} shows the angular size distribution function,
$f(\theta) \propto \theta^{-2.4}$, which peaks between $0.3 - 0.5$ deg$^2$.  
The median angular size of a cataloged cloud is 0.41 deg$^2$.
The distribution of total sky area covered as a function of cloud area,
is shown in Fig.~\ref{fig:coverage}. The majority of sky area coverage by
cataloged anomalous--velocity \HI\ is contributed by features between 0.5 $-$ 3
deg$^2$ in size.
 The total area of southern sky covered by high-velocity emission features
is 1582 deg$^2$.

The distribution of position angles of the anomalous objects with respect to
Galactic coordinates is shown in Fig.~\ref{fig:PAglact}, for those objects 
with a minor--to--major axial ratio $< 0.7$. The distribution is
relatively uniform, with the exception of an enhancement within about
20\deg~ of 0\deg/180\deg, implying an excess alignment perpendicular to the
Galactic plane for about 10\% of all cataloged objects.
This excess is largely the same population of clouds which follows
the general orientation of the Magellanic Stream.  Fig.~\ref{fig:PAms} shows the distribution of
position angles with respect to the Magellanic coordinate system, as
defined below.  The North pole of these coordinates is defined to lie
at Galactic coordinates $(l, b) = (180\deg, 0\deg)$; the equator of the
Magellanic system, i.e. latitude $B_{\rm M}$, forms a great circle
passing through the South Galactic pole and through the Galactic
equator at $l=90\deg~$ and 270\deg.  The Magellanic Stream lies roughly
parallel to the equator of the Magellanic coordinate system, but
generally slightly to its south.  The zero of Magellanic longitude,
$L_{\rm M}=0\deg$, is defined to coincide with the position of
the LMC.  The peak near 90\deg~ in Fig.~\ref{fig:PAms} shows an excess of about 10\% of
cataloged objects to be aligned parallel with the Magellanic Stream.



\subsection{Distribution of the compact high--velocity clouds}

If any of the anomalous--velocity clouds are at Local Group or larger
distances, they are likely to be small and distinct from the large
high--velocity complexes.  We classified 179 of the cataloged
objects as compact, isolated CHVCs, following the criteria outlined
in Section 2.4. A further 159 could not be unambiguously classified as CHVCs and
received the :HVC designation.  This number is well in excess of the 65
compact, isolated clouds cataloged by BB99 between Decl.s
$-30\deg~$ and 90\deg. Of the BB99 CHVCs, we find that about 50\% in the
Decl. range $-30\deg~$ to 2\deg~ do not satisfy the more stringent
selection criteria that we have adopted for a CHVC classification in
the HIPASS data. With the higher resolution and sensitivity we detected
elongations at low brightnesses in some cases which make their
association with nearby filamentary structures a possibility.

This section describes only the compact and isolated CHVCs cataloged in
the HIPASS data. The spatial distribution of the CHVCs is shown in the
polar--cap projection in Fig.~\ref{fig:Cdispersion}. The same local
minima noted above for the entire anomalous--velocity \HI distribution
apply here: due to survey incompleteness there is a strong
anti--correlation with the Galactic disk as well as with the Magellanic
Clouds.
Several apparent concentrations of CHVCs are seen in excess of a more
nearly uniform distribution. We will refer to the most prominent
concentrations by name: Group~1, extending from about
(RA, Decl.)$=(23^{\rm h}\rightarrow 2^{\rm h},
0^\circ\rightarrow-45^\circ)$; Group~2, from (RA, Decl.)$=(6^{\rm
h}\rightarrow 12^{\rm h}, -45^\circ\rightarrow-70^\circ)$; and Group~3,
from (RA, Decl.)$=(19^{\rm h}\rightarrow 21^{\rm h},
0^\circ\rightarrow-35^\circ)$.  It must be borne in mind, however, that
the appearance of such concentrations will be strongly modified by the
variations in survey completeness. In particular, Group~1 is
bisected by a bright portion of the Magellanic Stream, while Group~2 is
adjacent to a region of extensive Galactic emission, as can be seen by
comparison with Figs.~\ref{fig:hi_pos1} and \ref{fig:hi_neg}.
The distribution of the compact high-velocity clouds in Galactic
coordinates is shown in Fig. \ref{fig:chvcgal}.
In Galactic coordinates, the CHVC concentrations fall
near the South Galactic Pole (Group 1), near ($l,b$)~=~(270\deg,$-$15\deg) (Group 2),
and near ($l,b$)~=~(10\deg,$-$20\deg) (Group 3). 

The Group 1 CHVC concentration extends across the Magellanic Stream and
is also spatially co--extensive with the Sculptor Group of galaxies, though the most
prominent CHVC peak is offset from the center of both the Stream and the Sculptor Group.  
This group of CHVCs has an extreme variation in velocity over a limited 
region of the sky. We
have calculated the velocity dispersion as a function of position on the
sky for all CHVCs within a radius of $15^\circ$, whenever at least ten 
objects fall within this circle. The
resulting contours of velocity dispersion are drawn in
Fig.\ref{fig:Cdispersion}. Peak dispersions of 150~\kms~are found
toward Group~1. Though this value is affected by the $V_{\rm LSR}$ cutoff,
comparable velocity dispersions are seen nowhere else
in the Southern sky, nor in the CHVC compilation of BB99.
See Putman \etal\ (2002) and Putman (2000) for a thorough description of this
sightline through the Magellanic Stream.

Group~2 is spatially in the region leading
the Magellanic Clouds and is also near the Local Group
anti-barycenter position (RA, Decl.)$=(11^{\rm h}, -45^\circ)$,
where Blitz et al.~(1999) predict an excess of Local Group
debris.  The velocities of Group 2 are
predominantly positive in the {\sc lsr} frame.
Group~3 is the most diffuse of the CHVC concentrations, with only a
minor increase in the local number density. This concentration overlaps
with the region designated the GCN (Galactic Center Negative--velocity)
Complex by WvW91.

The angular two-point correlation function, $w(\theta)$, for CHVCs 
is significantly positive for separations on the sky,
$\theta < 20$\deg~ (Fig.~\ref{fig:autocorr}), confirming that CHVCs
are clustered.  $w(\theta)$ for all of the HVCs is also shown in 
Fig.~\ref{fig:autocorr}, and they are found to be less
clustered than the CHVCs.
$w(\theta)$ is computed by counting pairwise
separations, $N(\theta)$ in the range $(\theta, \theta+d\theta)$ and
evaluating

\begin{equation}
w(\theta)=\frac{\rho_r^2 N(\theta)}{\rho^2 N_r(\theta)} -1,
\label{eq:acfn}
\end{equation}
where $\rho$ is the mean cloud density, $N_r(\theta)$ is the
number of pairwise separations in a random cloud catalog, and
$\rho_r$ is the random cloud density.
The random ($n=10^4$) cloud catalog has identical
selection in Decl. ($<$ 2\deg) and excludes Galactic latitudes within
$|b|$ = 10\deg~ due to the velocity bias against this region
in the HVC catalog.  $w(\theta)$ gives the excess probability,
over the Poisson probability, of finding two clouds with separation
$\theta$. The random cloud catalog permits correlations arising from the
angular selection function to be effectively removed.  See Hamilton (1993) for a discussion of correlation functions and estimators.  For CHVCs, 
$w(\theta)$ appears to be approximated by a power
law with $w(\theta) = (\theta/10$\deg)$^{-0.6}$. For HVCs, the slope is
similar, but better-defined with $w(\theta) = (\theta/2$\deg$)^{-0.8}$.
CHVCs are significantly more clustered than HVCs. In the range
2\deg~ $< \theta <$ 20\deg, CHVCs have twice the excess probability
of being paired compared with HVCs.

The velocity distribution of the cataloged CHVCs is plotted against
Galactic longitude and latitude, for the Local, Galactic, and Local
Group reference frames in Figs.~\ref{fig:Cglon_vlsr} and
\ref{fig:Cvlsr_glat}, respectively.  The velocity histograms in
these reference frames are shown in Fig.~\ref{fig:Cvlsr_histo}.  
Comparison of the CHVC velocity distributions with those of all the clouds suggests that
the CHVCs are more equally distributed between positive and negative
{\sc lsr} velocities.  In the {\sc gsr} frame, the prominent peak 
at V$_{\rm GSR}=-30$ \kms~in Fig.~\ref{fig:vlsr_histo}
is not seen in the CHVC distribution shown in Fig.~\ref{fig:Cvlsr_histo}. 
The velocity distribution of the CHVCs declines from a dispersion
of 125~\kms~in the {\sc gsr} frame to 116~\kms~in the {\sc lgsr} frame.
The high--positive--velocity tail
in the {\sc lgsr} distribution of the CHVCs is contributed to a large
extent by the objects of Group~1.
Table 2 shows that the median {\sc lsr}, {\sc gsr}, and {\sc lgsr} velocities 
(95, $-38$ and $-57$ \kms, respectively)
are only slightly different from the other cloud populations.
The mean velocities are 12 \kms~({\sc lsr}), -47 \kms~({\sc gsr})
and -56 \kms~({\sc lgsr}).

The distributions of CHVC fluxes, peak column densities, and sizes are
shown in the log-log plots of Figs.~\ref{fig:Cflux_histo}, \ref{fig:Cnhi_histo}, and
\ref{fig:Csize_histo}.  Steep distributions are found with low end
turn--overs that are consistent with the survey completeness limits.
 Flattening of the flux
distribution from the form $f(F_{\rm HI}) \propto F_{\rm HI}^{-2.1}$ begins
at $\sim 30$~Jy~\kms, while the peak column--density distribution
begins to flatten from $f(N_{\rm HI}) \propto N_{\rm HI}^{-2.8}$ near
$2 \times 10^{19}$\,cm$^{-2}$. The peak in the size distribution,  $f(\theta) \propto \theta^{-2.5}$,
is at $\sim 0.3$ deg$^2$, which corresponds to significantly resolved
sources.
 Plotting total sky area covered as a function of cloud area
in Fig.~\ref{fig:Ccoverage}, reveals that most of the observed CHVC covering factor
is indeed due to resolved objects with a characteristic size
of about 0.5 deg$^2$.  The total sky area covered by the southern CHVCs
is 85 deg$^2$, with the largest contributors being CHVCs with sizes
between $0.4 - 1$ deg$^2$.

Table 2 shows that the CHVCs have a higher median 
total flux and peak column density than the entire anomalous--velocity cloud sample.
This may be largely due to the selection criteria for a CHVC requiring
a peak column density $> 12\sigma$. The median size for a CHVC is 0.36
deg$^2$, which is only slightly smaller than the median size of a cataloged
HVC (0.42 deg$^2$), despite the 2\arcdeg\ upper limit
to angular size in their selection.  Therefore, the main distinguishing
feature of HIPASS CHVCs is isolation, rather than compactness.
 The properties of the CHVCs were investigated in terms of
Galactic coordinates and no  
gradient in individual total flux with position was found.
This also holds for the peak column densities and sizes.  
By definition, the CHVCs do not have a preference for a specific position
angle.  The :HVCs also do not have a preferred position angle.

 
\section{Discussion \& Summary}

The high resolution and sensitivity of HIPASS, combined with
repeated coverages that together yield Nyquist sampling and
source confirmation, have provided an unprecedented view of
anomalous--velocity \HI emission in the Southern hemisphere. These
attributes make it possible to detect the morphology of emission
features and utilize this in their classification, rather than relying
only on their position and velocity. The majority of emission features
are associated with relatively diffuse filaments with modest
variations in surface brightness.
In addition to an assortment of filamentary complexes, there is also
a distinct population of compact, isolated objects with a high column
density contrast, the CHVCs.  The superior resolution and sensitivity
of the HIPASS data have allowed much better discrimination of CHVC and
HVC sources than was possible previously.

\subsection{HVCs}

As is evident from a comparison of Figs.~\ref{fig:hi_pos1} to
\ref{fig:hi_neg} with Fig.~\ref{fig:ra_dec}, the Southern sky is rich
in anomalous--velocity \HI. The total flux of these objects is
dominated by the Magellanic Stream, which accounts for approximately $2.1
\times 10^8$ \Msun\ of neutral hydrogen (at an average distance of
55~kpc) and is discussed in detail by Putman et
al. (2002) and Putman (2000). Also evident from these figures is the perceived
avoidance by the anomalous--velocity objects of the Galactic disk and
the Magellanic Clouds, largely due to a variable degree of survey
completeness resulting from confusion with regions of bright
emission extending over a wide range of velocities.
One of the most striking features of the spatial distribution is the
paucity of anomalous--velocity \HI in a region (unaffected by variable
completeness) bounded approximately by (RA, Decl.)$=(2^{\rm
h}\rightarrow 5^{\rm h}, -10^\circ\rightarrow-30^\circ)$.
In Galactic coordinates, this is centered at ($\ell, b$) $\approx (200\deg,
-50\deg)$, which is both well off the Galactic plane and along a sightline
which extends over only one spiral arm (see Taylor \& Cordes 1993).  The
percentage of our Galaxy which lies along a particular sightline may have
implications on the number of HVCs in that direction.

The distributions of total flux and peak column density for the entire
population of cataloged anomalous velocity clouds can be
described by steep power laws (of slopes $-2.1$ and $-2.9$,
respectively) with a flattening at low values due to
the nominal survey completeness limit. The
distribution of apparent cloud sizes is strongly peaked at angular sizes in the
range $0.3 - 0.5$ deg$^2$.  These properties imply that the
majority of cataloged objects share a characteristic flux, $\sim20$~Jy~\kms;
peak column density, $\sim10^{19}$ cm$^{-2}$; and size, $\sim 0.4$ deg$^2$,
in agreement with the median properties listed in Table 2.
Unless the clouds can be assigned with a 
characteristic distance, the distribution
of total flux can not be easily converted into an \HI mass function of
anomalous--velocity gas.  If one assumes a characteristic distance, D,
the mass of a typical cloud is $\sim$ 4.5 D$_{\rm kpc}^2$ \Msun.

The distribution of position angles of the high--velocity clouds
indicates that an excess of about 10\% are aligned along the long axis
of the Magellanic Stream and may have a common origin. A comparable
excess of objects is found elongated perpendicular to the Galactic
plane, although this is basically the same population of
objects, given the orientation of the Stream.
This excess remains when the narrow region of sky traditionally
associated with the Magellanic Stream is excluded from the analysis.
 The elongation of the clouds may represent an interaction with
material in the Galactic halo or the strength of the  
gravitational field of the Milky Way.  Quilis \& Moore (2001)
find that head-tail features in HVCs (similar to the :HVC pictured in
Fig.~\ref{fig:grouping}) can be reproduced in either pure gas clouds, or clouds
embedded in dark matter halos, as long as the Galactic wind density is higher 
than
10$^{-4}$ cm$^{-3}$.  They suggest that all HVCs will show tails with higher 
sensitivity observations.
The reclassification of many of the BB99
CHVCs due to detection of low column density extensions from the cloud cores 
with HIPASS, may represent the detection of these tails.

Based on the completeness levels of the catalog (see section 2.5), high-velocity 
features
cover 19\% of the southern sky.  This is almost exactly the value found by
Wakker (1991) for $|V_{\rm LSR}| >  100$ \kms~and N$_{\rm HI} > 2 \times
10^{18}$ cm$^{-2}$
using the WvW91 all-sky catalog.  Murphy, Lockman \& Savage (1995) found a 
covering
fraction of 37\% in the regions around AGNs for N$_{\rm HI} > 7 \times
10^{17}$ cm$^{-2}$.  This covering fraction agrees with the extrapolation of 
N$_{\rm
HI}$
down to this level in the all-sky surveys (Wakker 1991).  The
detection of high-velocity O VI and Mg II absorption lines (Sembach et al. 2000; 
Gibson et
al. 2000) and H$\alpha$ emission (Tufte et al. 1998) off the 21-cm survey 
contours also suggests a larger all-sky covering fraction for high velocity gas.   
The high-velocity cloud features
presented here
may simply be the high HI column density peaks among a dense web of multi-phase
halo material.    

The peak column--density distribution of the
anomalous--velocity clouds is significantly steeper than
 the distribution of column densities observed in quasar absorption lines 
(Penton et al. 2000; Hu et al. 1995) and (for N$_{\rm HI} < 10^{21}$
cm$^{-2}$) the galaxies of the Ursa Major cluster
(Zwaan, Verheijen, \& Briggs 1999).  These comparisons 
may be premature since we are sampling the distribution of
peak cloud column densities, rather than the distribution of
all column densities.
One would have to assume that the decreasing
radial distribution, which will dominate the surface area and hence the
statistics for random sight lines, behaves similarly for all clouds. 
Steps have been taken recently to
quantify the radial distributions of column density at the edges of
CHVCs (Burton, Braun, \& Chengalur 2001). Approximately exponential edge
profiles have been observed for $N_{\rm HI}$ between a few
$10^{17}$ and about $10^{19}$ cm$^{-2}$, although the scale lengths
vary among objects by at least a factor of five. 
Also, high resolution studies of HVCs (e.g. Wakker \&
Schwarz 1991, Braun \& Burton 2000) have shown great variation in 
column densities within cloud sub--structures,
down to at least arcmin scales. In practice, the variation in
apparent peak column density observed with 15\farcm5 versus $1'$ resolution
amounts to a factor of about 3 for the HVC complexes A, L, and M, and
ranges between a factor of 5 and 20 for the CHVCs which have been
studied to date.

Some parallels can be made between those HVCs not directly associated with the
Magellanic System (e.g. not the Stream, Leading Arm, and Bridge) and the 
anomalous velocity gas seen in NGC 2403 by Sancisi et al. (2001).  
The anomalous velocity components of NGC 2403 make up 10\% of its total
\HI\ 
mass, with individual long filaments having masses of $\sim 10^6$ and $10^7$ \Msun.  
At a mean distance of 15 kpc, the non-Magellanic HVCs would be $\sim10$\% of the Galaxy's \HI\ mass, and the mass
estimates for some of the large high-velocity complexes with distance constraints are $\sim 10^6$ \Msun~(Complexes A and C; van Woerden et al.
1999; Wakker et al. 1999).  Further parallels can be made when one considers that some low $V_{\rm GSR}$ HVCs show clear links to Galactic \HI\ when examined in both position and velocity (e.g. figure 6 in Putman \& Gibson 1999), and much of the extreme positive and negative high-velocity gas can be associated with the Magellanic HVCs.  
The XHVCs listed in Table 3 merge with Galactic emission, and though some
may simply lie at overlapping velocities, they have a higher likelihood of being 
directly related to the Galaxy.  Further deep \HI\ studies of Milky
Way-like spirals will provide important insight into the nature of the
anomalous velocity gas which surrounds our Galaxy.

\subsection{CHVCs}

We have cataloged a substantial population (179) of compact,
isolated high--velocity \HI clouds (CHVCs) in the Southern hemisphere as shown
in Fig. \ref{fig:Cdispersion}. The quality of the HIPASS data has
resulted in this sample being larger than the sample of
65 objects north of Decl. $-30^\circ$ cataloged by BB99, despite
the fact that we have applied more stringent selection criterion. For
our HIPASS sample we have demanded that the 25\% column--density
contour of each object (1) be closed, with a diameter less than
2$^\circ$; (2) not be elongated in the direction of any nearby extended
emission; and (3) be well above the local noise floor (implying a peak
signal--to--noise ratio of at least 12$\sigma$). By comparison, BB99
applied the same three criteria to the 50\% column density contour seen
in the LDS data. Only 50\% of the BB99 CHVC sample in the Decl.
zone of overlap satisfy these more demanding conditions in the HIPASS
data.  These results suggest that the CHVC classification is dependent
on the sensitivity and resolution of the survey used to classify them;
however, the increase in the number of catalogued CHVCs could also
partially be due to a bias in their overall distribution for
the Southern hemisphere.
In a subsequent paper (de Heij et al. 2001) the
Leiden/Dwingeloo database is investigated with the same search algorithm 
and selection criteria employed here.

The distribution of cataloged Southern CHVCs is not uniform on the
sky.  The angular two--point correlation function of CHVCs, shown in 
Fig. \ref{fig:autocorr}, suggests that they are approximately twice as
clustered as the more general HVC population, out to angular
scales of about 20$^\circ$.  There are several possible reasons
for the clustering.  It may be partially
a selection effect due to diminished sensitivity in certain regions
of the sky (as discussed in section 2.5). 
The clustering could also reflect the formation or destruction process
of a larger cloud.  Finally, CHVCs could represent high column density
peaks within a more diffuse medium, similar to what is proposed
for galaxies and the Ly$\alpha$ absorber systems which also appear
clustered (Dave et al. 1999).
This scenario could also explain the diffuse tails seen around many
of the CHVCs and :HVCs (see also the results of Br\"uns et al. 2000 and Burton et al. 2001).  However,
as stated in the previous section, the elongation of these clouds could also represent interaction
with the Galaxy's halo.

The CHVCs cluster in several concentrations, which we have designated Groups~1, 2, and 3. 
Group~1 is spatially co--extensive with a portion of the Magellanic 
Stream and the Sculptor Group of galaxies near the
South Galactic pole (see Putman et al. 2002 and Putman 2000).  Kinematically, the distribution extends continuously from the
systemic velocity of Sculptor, near $V_{\rm GSR}=+200$ \kms, to high
negative velocities, $V_{\rm GSR}=-300$ \kms, but it is unknown
what the distribution does between
$-60<V_{\rm dev}<60$ \kms, where there is confusion with Galactic emission. 
The Magellanic Stream is on the negative velocity side of this undetectable 
range in this part of the sky.
Since this concentration lies near the Galactic pole, the
corresponding velocities in the {\sc lsr}, {\sc gsr}, and {\sc lgsr}
frames are virtually identical.   This kinematic range of 500~\kms~is
responsible for a local peak in the velocity dispersion of CHVC
objects, which is not seen anywhere else in the Southern sky (see 
Fig. \ref{fig:Cdispersion}).  The
Group~2 concentration of CHVCs is a more diffuse collection of objects
with high positive {\sc lsr} velocities (which have a mean near 0
\kms~in both the {\sc gsr} and {\sc lgsr} frames). As noted previously,
this concentration is in the Local Group anti-barycenter region and also the
area leading the Magellanic Clouds; however, it also lies next to
the Galactic plane, so the apparent distribution may be
strongly affected by confusion effects. A third diffuse concentration
of CHVCs, Group~3, is characterized by high negative velocities in the
both the {\sc lsr} and {\sc gsr} frames.


If one presumes that CHVCs are self-gravitating gas clouds, their distance can be estimated from the virial theorem
using,
\begin{equation} 
D_{vir} = f \frac{3 \, \sigma^2 \sqrt{\Omega}} {2 \sqrt{\pi}  \, G  \, 0.236  \, F_{HI}} 
\end{equation}
(following Wakker et al. (2001)). $f$ is M(HI)$/$M(total), $\sigma$ is the velocity dispersion in units
of kpc Myr$^{-2}$ ($= \sigma$ (\kms) $\times\ 1.023 \times 10^{-3}$), 
$\Omega$ is the solid angle of the CHVC, $G = 4.53 \times 10^{-12}$ kpc$^3$ Myr$^{-2}$ \Msun$^{-1}$, and $F_{\rm HI}$ is
the \HI\ flux in Jy~\kms.
 The median angular size, $\Omega$, of a HIPASS CHVC is 0.36 deg$^2$, the velocity
dispersion, $\sigma$, is typically 10-15 \kms,
and the median total flux, $F_{\rm HI}$, is 19.9 Jy \kms.  
If $f=0.1$, the CHVCs are at a median distance, $D_{vir}$, of 400 kpc. 
With an estimate of a CHVC's distance, the total flux can be thought
of in terms
of its \HI\ mass using $M_{\rm HI} = 0.236~F_{\rm HI}$ (Jy~\kms) $D^2_{\rm kpc}$ \Msun.  
At a distance of 400 kpc, the entire population of CHVCs has a range of
\HI\ masses from 3.8 $\times 10^4$ \Msun\ to 3.4 $\times 10^7$ \Msun, with a median value of 7.5 $\times 10^5$ \Msun. 
Moving the clouds out to 700 kpc ($f=0.3$) gives a median \HI\ mass of 2.3
$\times 10^6$ \Msun\
and 55 kpc ($f=0.02 $) gives $1.4 \times 10^4$ \Msun.
The actual value of M(HI)$/$M(total) and whether or not CHVCs are self-gravitating
remains to be seen.  If CHVCs are not self-gravitating and there is no
other confining medium, the clouds are unlikely to last more than a few
hundred Myrs (again depending on the cloud's distance; e.g. see Blitz et
al. 1999).  Using the classic crossing time argument, at 55 kpc, a typical CHVC would double in size in only 25 Myr.

The distribution of CHVC sizes has a slightly steeper slope than 
the entire population of clouds, probably because of the 2\arcdeg\ upper size
limit imposed in the catalog.  CHVCs cover only 1\% of the southern sky.
The CHVC distributions of flux and peak column density have approximately
the same slope as the entire population of HVCs.  So again, the N$_{HI}$
slope is steeper than that of a sample of galaxies.  The $F_{\rm HI}$ slope
is actually  shallower than a that of a sample of galaxies taken from
the HIPASS survey  ($f(F_{\rm HI}) \propto F_{\rm HI}^{-2.7}$; Kilborn 2000; Putman 2000).  There is no systematic
distribution of cloud fluxes and sizes with velocity or position which
could be used to argue for a greater distance in certain regions of the
sky or at higher velocities.  
CHVCs have a median {\rm GSR} velocity of -38 \kms, which is suggestive 
of infall and is similar to the Northern population of CHVCs (BB99).
The median and mean {\rm LGSR} velocity is slightly lower for the CHVCs
than the other populations of clouds.

\section{Acknowledgements} 

  We are grateful to Mark Calabretta for his help in developing
the software for reducing HIPASS.  M. Putman thanks the
Australia Telescope National Facility for hosting her during
much of this work, and Zonta International for travel funding.  Swinburne 
University is thanked for the use of their supercomputing facility.
The Netherlands
Foundation for Research in Astronomy is operated under contract with
The Netherlands Organization for Scientific Research.  This research
has made use of the NASA/IPAC Extragalactic Database (NED) which is
operated by the Jet Propulsion Laboratory, California Institute of
Technology, under contract with the National Aeronautics and Space
Administration.  We have also made use of the LEDA database
(http://leda.univ-lyon1.fr).  The Digitized Sky Surveys were produced
at the Space Telescope Science Institute under U.S. Government grant
NAG W-2166. 


\begin{figure*} 
\caption{Representative column density images of two areas of the sky. The
\HI emission centered on the cloud HVC$245.3-79.5-096$ is shown on the
left, integrated over the velocity range from $-73$ \kms~to $-152$ \kms.
HVC$012.4-60.5-136$ is shown on the right, integrated from $-86$ \kms~to
$-152$ \kms.  The grey--scale bar indicates the intensity scaling in
units of K\,\kms.  Each cloud is indicated by a box which encloses all
pixels identified with the cloud.  All of the clouds cataloged from
these two areas of the sky and found to have some emission within the
velocity ranges are indicated by boxes, including those centered at
velocities outside the integration ranges represented in these two
panels. Those boxes which appear empty indicate clouds which only have
their velocity extremes within the image.  The more extended clouds in
the left-hand image are not cataloged because they merge with emission
at the conventional velocities associated with the Milky Way.  }
\label{fig:examples}
\end{figure*}

\begin{figure*} 
\caption{Illustration of cloud classification. The integrated \HI
emission over the velocity extent of each of the three objects is shown in a
$10^\circ \times 10^\circ$ field. The grey--scale wedge indicates the intensity
scaling in units of K\,\kms. Contours are drawn at 30, 50, 70, and 90\%
of the peak column density. The heavy bar in the lower left corner
indicates a length of 2$^\circ$. The object on the left has been given
the CHVC designation, based on the absence of elongation of the 30\%
contour along filamentary structure in the field. The object on the
right has been classified as HVC, based on the clear extension of the
30\% contour. The object in the center panel has been given the :HVC
designation since there is some ambiguity in its elongation.}
\label{fig:grouping}
\end{figure*}

\begin{figure*} 
\caption{Integrated--intensity map of the anomalous--velocity HIPASS
sky in a South celestial pole projection showing emission at
positive velocities ranging from $V_{\rm LSR} = 99$ to $500$ \kms.  The
Decl. circles are in increments of 30\deg.  The
Galactic equator and the great circle at $l = 0^\circ$ are indicated with
solid lines, while the South Galactic pole is indicated with an open
circle.  The grey-scale bar on the left indicates the intensity scale in
units of K \kms.  The Large and Small Magellanic Clouds (LMC and SMC) and the Galactic plane
dominate the intensities in this velocity range and much of their emission
is saturated in this grey--scale image. The Magellanic Bridge (between the LMC and SMC), the portion of the Magellanic
Stream in this velocity range (between the Clouds and the South Galactic pole), and the Leading Arm (between the Clouds and
the Galactic plane) are also shown. The
Galactic plane shows some effect of the emission filling
the entire 8\deg~ scan and being affected by the bandpass correction
(see Putman et al. 2001). }
\label{fig:hi_pos1}
\end{figure*}

\begin{figure*} 
\caption{Integrated--intensity map of the anomalous--velocity HIPASS
sky at more extreme positive velocities than Fig. 3.  This map includes
$V_{\rm LSR} = 200$ to 500 \kms, a range which excludes most of the emission from the
Galactic plane.  The LMC, the Magellanic Bridge, and the Leading Arm
dominate the \HI flux observed, but numerous other HVC and CHVC features
are evident.   }
\label{fig:hi_pos2}
\end{figure*}

\begin{figure*} 
\caption{Integrated--intensity map of the anomalous--velocity HIPASS
sky at negative velocities extending from $V_{\rm LSR} = -99$ to $-500$
\kms~ in the same projection as Fig. 3.  Portions of the Magellanic Stream and of the Galactic plane
dominate the emission at these velocities, and are indicated. }
\label{fig:hi_neg}
\end{figure*}

\begin{figure*} 
\caption{Distribution of all anomalous--velocity objects cataloged (HVCs,
CHVCs and :HVCs) in
the south--polar--cap projection.  The
Decl. circles are in increments of 30\deg.
No distinction is made here between
the extended HVCs and the compact CHVCs. Plus and minus symbols
are used to indicate positive and negative velocities in the {\sc gsr} frame,
with symbol size proportional to magnitude of their {\sc gsr} velocity.  }
\label{fig:ra_dec}
\end{figure*}

\begin{figure*} 
\caption{Distribution of all anomalous--velocity objects cataloged in
Galactic coordinates. $l = 0$\deg~is in the center of this figure, and
increases in increments of 45\deg~to the left.  $b$ increases in increments
of 30\deg.}
\label{fig:gal}
\end{figure*}

\begin{figure} 
\caption{Velocities of all of the anomalous--velocity objects
identified in the HIPASS data, plotted against Galactic longitude for
three different kinematic reference frames.  The
motions of the objects are shown measured with respect to the Local
Standard of Rest (a); with respect to the Galactic
Standard of Rest (b); and with respect to the Local
Group Standard of Rest (c). The bottom panel shows the number of
high--velocity clouds in terms of Galactic Longitude.}
\label{fig:glon_vlsr}
\end{figure}

\begin{figure} 
\caption{Galactic Latitude versus velocity for all of the
anomalous--velocity objects identified in the HIPASS data, in three
different kinematic reference frames.  In the upper panel, the motions
of the objects are shown measured with respect to the Local Standard
of Rest; in the middle panel, with respect to the Galactic Standard of
Rest; and in the lower panel, with respect to the Local Group Standard
of Rest. }
\label{fig:vlsr_glat}
\end{figure}

\begin{figure} 
\caption{Histogram of velocities of all of the
anomalous--velocity objects identified in the HIPASS data, in three
different kinematic reference frames.  In the upper panel, the motions
of the objects are shown measured with respect to the Local Standard
of Rest; in the middle panel, with respect to the Galactic Standard of
Rest; and in the lower panel, with respect to the Local Group Standard
of Rest. }
\label{fig:vlsr_histo}
\end{figure}

\begin{figure} 
\caption{ Distribution of total \HI\ fluxes, $F_{\rm
HI}$, for all of the anomalous--velocity objects cataloged.  The data
are binned in equal intervals of $\log(F_{\rm HI})$. The slope of
$-1.1$ implies a distribution function, in linear units, $f(F_{\rm HI})
\propto F_{\rm HI}^{-2.1}$.  }
\label{fig:flux_histo}
\end{figure}

\begin{figure} 
\caption{ Distribution of peak \HI column density, $N_{\rm
HI}$, for all of the anomalous--velocity objects cataloged.  The data
are binned in equal intervals of $\log($N$_{\rm HI})$. The slope of
$-1.9$ implies a distribution function, in linear units, $f($N$_{\rm HI})
\propto $N$_{\rm HI}^{-2.9}$.  }
\label{fig:nhi_histo}
\end{figure}

\begin{figure} 
\caption{Distribution of angular size, $\theta$, for
all of the anomalous--velocity objects cataloged.  The data are binned
in equal intervals of $\log(\theta)$. The distribution is highly peaked
between 0.3 deg$^{2}$ and 0.5 deg$^{2}$ and implies a distribution function, in linear
units, $f(\theta) \propto \theta^{-2.4}$.}
\label{fig:size_histo}
\end{figure}

\begin{figure} 
\caption{Distribution of total sky area covered as a function of cloud
area for all of the anomalous--velocity objects cataloged. The largest
contribution to areal coverage is provided by objects with an angular
size of $ \sim 2$ deg$^2$ in angular size.
 The total area of southern sky covered by high-velocity emission features
is 1582 deg$^{2}$. }
\label{fig:coverage}
\end{figure}

\begin{figure} 
\caption{Distribution of position angles, with respect to
Galactic coordinates, for all of the anomalous--velocity objects
cataloged.}
\label{fig:PAglact}
\end{figure}

\begin{figure} 
\caption{Distribution of position angles, with respect
to Magellanic coordinates, for all of the anomalous--velocity objects cataloged.  }
\label{fig:PAms}
\end{figure}

\begin{figure*} 
\caption{Distribution of CHVC objects cataloged in
the south--polar--cap projection.  Plus and minus symbols
are used to indicate positive and negative velocities in the {\sc gsr} frame,
with symbol size proportional to magnitude. Contours of local velocity
dispersion are drawn at values of 50, 100, and 150~\kms. }
\label{fig:Cdispersion}
\end{figure*}

\begin{figure*}
\caption{Distribution of the compact high-velocity clouds (CHVCs) in Galactic coordinates.}
\label{fig:chvcgal}
\end{figure*}

\clearpage
\begin{figure} 
\caption{The angular two-point correlation function, $w(\theta)$,
for HVCs (triangles) and CHVCs (circles).}
\label{fig:autocorr}
\end{figure}

\begin{figure} 
\caption{Velocities of CHVC objects
identified in the HIPASS data, plotted against Galactic longitude for
three different kinematic reference frames.  The
motions of the CHVCs are shown measured with respect to the Local
Standard of Rest (a); with respect to the Galactic
Standard of Rest (b); and with respect to the Local
Group Standard of Rest (c). The bottom panel shows the number of
compact high--velocity clouds in terms of Galactic Longitude. }
\label{fig:Cglon_vlsr}
\end{figure}

\begin{figure} 
\caption{Galactic latitude versus velocities of CHVC objects
identified in the HIPASS data, in three different kinematic reference
frames.  In the upper panel, the motions of the objects are shown
measured with respect to the Local Standard of Rest; in the middle
panel, with respect to the Galactic Standard of Rest; and in the lower
panel, with respect to the Local Group Standard of Rest. }
\label{fig:Cvlsr_glat}
\end{figure}

\begin{figure} 
\caption{Histogram of velocities of CHVC objects identified in the
HIPASS data, in three different kinematic reference frames.  In the
upper panel, the motions of the objects are shown measured with
respect to the Local Standard of Rest; in the middle panel, with
respect to the Galactic Standard of Rest; and in the lower panel, with
respect to the Local Group Standard of Rest. }
\label{fig:Cvlsr_histo}
\end{figure}

\begin{figure} 
\caption{ Distribution of total \HI fluxes, $F_{\rm
HI}$, for the CHVCs cataloged.  The data
are binned in equal intervals of $\log(F_{\rm HI})$. The slope of
$-1.1$ implies a distribution function, in linear units, $f(F_{\rm HI})
\propto F_{\rm HI}^{-2.1}$. }
\label{fig:Cflux_histo}
\end{figure}

\begin{figure} 
\caption{ Distribution of peak \HI column density, $N_{\rm
HI}$, for the CHVCs cataloged.  The data
are binned in equal intervals of $\log($N$_{\rm HI})$. The slope of
$-1.8$ implies a distribution function, in linear units, $f($N$_{\rm HI})
\propto $N$_{\rm HI}^{-2.8}$.  }
\label{fig:Cnhi_histo}
\end{figure}

\begin{figure}
\caption{Distribution of angular size, $\theta$, for the CHVCs cataloged.  The data are binned
in equal intervals of $\log(\theta)$. The distribution is highly peaked
between $0.3 - 0.4$ deg$^{2}$ and implies a distribution function, in linear
units, $f(\theta) \propto \theta^{-2.5}$ ($\theta^{-2.2}$ if the last two points
are excluded from the fit). }
\label{fig:Csize_histo}
\end{figure}

\begin{figure} 
\caption{Distribution of total sky area covered as a function of cloud
area for the CHVCs cataloged. The total sky area covered by the southern CHVCs
is 85 deg$^2$. }
\label{fig:Ccoverage}
\end{figure}

\clearpage
\setlength{\baselineskip}{8pt}
\font\cmreight=cmr7
\setlength{\tabcolsep}{1mm}
\renewcommand{\arraystretch}{0.6}
\newcommand{\mtm}{$\mtf -$}

\newcommand{\tableallign}{|c||rl||rrrrrr||rrrr||rrr||r|}
\newcommand{\tableheader}{
	\multicolumn{1}{|c||}{\tf \#}            &
	\multicolumn{2}{c||}{\tf designation}    &
	\multicolumn{1}{c}{\tf RA (J2000)}       &
	\multicolumn{1}{c}{\tf DEC (J2000)}      &
	\multicolumn{1}{c}{$\mtf v_{\rm LSR}$}   &
	\multicolumn{1}{c}{$\mtf v_{\rm GSR}$}   &
	\multicolumn{1}{c}{$\mtf v_{\rm LGSR}$}  &
	\multicolumn{1}{c||}{\tf FWHM}           &
	\multicolumn{1}{c}{\tf MAJ}              &
	\multicolumn{1}{c}{\tf MIN}              &
	\multicolumn{1}{c}{\tf PA}               &
	\multicolumn{1}{c||}{\tf size}           &
	\multicolumn{1}{c}{$\mtf T_{\rm peak}$}  &
	\multicolumn{1}{c}{$\mtf N_{\rm HI}$}    &
	\multicolumn{1}{c||}{\tf flux}           & 
	\multicolumn{1}{c|}{\tf galaxy}         \\
	\multicolumn{1}{|c||}{}                           &
	\multicolumn{2}{c||}{$\mtf ll\pm bb\pm vvv$}      &
	\multicolumn{1}{c}{$\mtf (h\ m)$}                 &
	\multicolumn{1}{c}{$\mtf (\circ\ \prime)$}        &
	\multicolumn{1}{c}{$\mtf (\rm km\;s^{-1})$}       &
	\multicolumn{1}{c}{$\mtf (\rm km\;s^{-1})$}       &
	\multicolumn{1}{c}{$\mtf (\rm km\;s^{-1})$}       &
	\multicolumn{1}{c||}{$\mtf (\rm km\;s^{-1})$}     &
	\multicolumn{1}{c}{$\mtf (\circ)$}                &
	\multicolumn{1}{c}{$\mtf (\circ)$}                &
	\multicolumn{1}{c}{$\mtf (\circ)$}                &
	\multicolumn{1}{c||}{$\mtf (\circ^2)$}            &
	\multicolumn{1}{c}{$\mtf (\rm K)$}                &
	\multicolumn{1}{c}{$\mtf (10^{20}\rm\;cm^{-2})$}  &
	\multicolumn{1}{c||}{$\mtf (\rm Jy\;km\;s^{-1})$} &
	\multicolumn{1}{c|}{}                            \\
  }

\begin{table}

\caption[]{The beginning of the HIPASS catalog of anomolous velocity
objects ($-500$ \kms $< V_{\rm LSR} < 500$ \kms;  Decl. $< +2$\deg).
A complete version of the catalog is available at {\it
http://wwwatnf.atnf.csiro.au/research/multibeam} and the electronic version of this journal. }

\begin{sideways}
\begin{tabular}{\tableallign}
\hline
\tableheader
\hline
{ \tf    1} & { \tf  HVC } & {\tf  000.0\mtm59.4\mtm114} & { \tf 22 36.5} & { \tf \mtm39 42} & { \tf \mtm114} & { \tf \mtm114} & { \tf \mtm115} & { \tf    40} & { \tf  0.3} & { \tf  0.2} & { \tf    40} & { \tf   0.11} & { \tf  0.05} & { \tf  0.04} & { \tf      5.9}  & \\ 
{ \tf    2} & { \tf  HVC } & {\tf  000.2\mtm11.5\mtm233} & { \tf 18 33.7} & { \tf \mtm34 13} & { \tf \mtm233} & { \tf \mtm232} & { \tf \mtm286} & { \tf    36} & { \tf  0.1} & { \tf  0.1} & { \tf    60} & { \tf   0.12} & { \tf  0.11} & { \tf  0.07} & { \tf      7.0}  & \\ 
{ \tf    3} & { \tf    HVC } & {\tf  000.4\mtm82.5+109} & { \tf 00 21.9} & { \tf \mtm30 58} & { \tf    109} & { \tf    110} & { \tf    136} & { \tf    38} & { \tf  0.3} & { \tf  0.2} & { \tf    20} & { \tf   1.19} & { \tf  0.15} & { \tf  0.12} & { \tf     91.2}  & \\ 
{ \tf    4} & { \tf      HVC } & {\tf  000.4+06.7+122} & { \tf 17 21.1} & { \tf \mtm24 59} & { \tf    122} & { \tf    124} & { \tf     58} & { \tf    46} & { \tf  0.2} & { \tf  0.2} & { \tf \mtm40} & { \tf   0.14} & { \tf  0.09} & { \tf  0.07} & { \tf      5.8}  & \\ 
{ \tf    5} & { \tf    HVC } & {\tf  000.5\mtm75.8+173} & { \tf 23 53.4} & { \tf \mtm34 01} & { \tf    173} & { \tf    174} & { \tf    192} & { \tf    50} & { \tf  0.2} & { \tf  0.2} & { \tf \mtm40} & { \tf   0.44} & { \tf  0.16} & { \tf  0.16} & { \tf     32.3}  & \\ 
{ \tf    6} & { \tf   CHVC } & {\tf  000.6+21.3\mtm104} & { \tf 16 31.6} & { \tf \mtm16 08} & { \tf \mtm104} & { \tf \mtm102} & { \tf \mtm172} & { \tf    39} & { \tf  0.2} & { \tf  0.2} & { \tf \mtm10} & { \tf   0.32} & { \tf  0.21} & { \tf  0.16} & { \tf     17.8}  & \\ 
{ \tf    7} & { \tf    HVC } & {\tf  000.9\mtm68.0+126} & { \tf 23 17.8} & { \tf \mtm36 51} & { \tf    126} & { \tf    127} & { \tf    136} & { \tf    37} & { \tf  0.2} & { \tf  0.1} & { \tf    70} & { \tf   0.07} & { \tf  0.06} & { \tf  0.04} & { \tf      3.7}  & \\ 
{ \tf    8} & { \tf  HVC } & {\tf  000.9\mtm54.8\mtm094} & { \tf 22 12.4} & { \tf \mtm40 08} & { \tf  \mtm94} & { \tf  \mtm92} & { \tf  \mtm99} & { \tf    77} & { \tf  0.3} & { \tf  0.2} & { \tf    10} & { \tf   0.37} & { \tf  0.12} & { \tf  0.18} & { \tf     23.5}  & \\ 
{ \tf    9} & { \tf    HVC } & {\tf  001.0+43.7\mtm152} & { \tf 15 22.8} & { \tf \mtm01 25} & { \tf \mtm152} & { \tf \mtm150} & { \tf \mtm218} & { \tf    39} & { \tf  0.2} & { \tf  0.2} & { \tf    40} & { \tf   0.07} & { \tf  0.05} & { \tf  0.03} & { \tf      3.2}  & \\ 
{ \tf   10} & { \tf :HVC } & {\tf  001.1\mtm66.4\mtm093} & { \tf 23 10.1} & { \tf \mtm37 16} & { \tf  \mtm93} & { \tf  \mtm91} & { \tf  \mtm84} & { \tf    34} & { \tf  0.2} & { \tf  0.1} & { \tf \mtm20} & { \tf   0.08} & { \tf  0.07} & { \tf  0.05} & { \tf      3.2}  & \\ 
{ \tf   11} & { \tf    HVC } & {\tf  001.1+06.1\mtm264} & { \tf 17 25.2} & { \tf \mtm24 46} & { \tf \mtm264} & { \tf \mtm259} & { \tf \mtm324} & { \tf    50} & { \tf  0.4} & { \tf  0.2} & { \tf    80} & { \tf   0.47} & { \tf  0.13} & { \tf  0.11} & { \tf     28.0}  & \\ 
{ \tf   12} & { \tf CHVC } & {\tf  001.2\mtm15.5\mtm186} & { \tf 18 53.5} & { \tf \mtm34 57} & { \tf \mtm186} & { \tf \mtm181} & { \tf \mtm231} & { \tf    34} & { \tf  0.4} & { \tf  0.2} & { \tf \mtm20} & { \tf   1.25} & { \tf  0.81} & { \tf  0.54} & { \tf    158.4}  & \\ 
{ \tf   13} & { \tf   :HVC } & {\tf  001.5\mtm67.3+122} & { \tf 23 14.2} & { \tf \mtm36 52} & { \tf    122} & { \tf    124} & { \tf    133} & { \tf    32} & { \tf  0.2} & { \tf  0.1} & { \tf    20} & { \tf   0.17} & { \tf  0.14} & { \tf  0.09} & { \tf      9.6}  & \\ 
{ \tf   14} & { \tf  HVC } & {\tf  001.6\mtm11.9\mtm213} & { \tf 18 38.2} & { \tf \mtm33 08} & { \tf \mtm213} & { \tf \mtm207} & { \tf \mtm259} & { \tf    38} & { \tf  1.2} & { \tf  0.6} & { \tf    40} & { \tf   0.07} & { \tf  0.06} & { \tf  0.04} & { \tf      2.5}  & \\ 
{ \tf   15} & { \tf  HVC } & {\tf  001.6\mtm08.9\mtm182} & { \tf 18 25.2} & { \tf \mtm31 50} & { \tf \mtm182} & { \tf \mtm176} & { \tf \mtm231} & { \tf    90} & { \tf  0.3} & { \tf  0.3} & { \tf    50} & { \tf   0.52} & { \tf  0.07} & { \tf  0.09} & { \tf     24.9}  & \\ 
{ \tf   16} & { \tf    HVC } & {\tf  001.6+43.1\mtm161} & { \tf 15 25.7} & { \tf \mtm01 31} & { \tf \mtm161} & { \tf \mtm157} & { \tf \mtm225} & { \tf    32} & { \tf  0.2} & { \tf  0.2} & { \tf    20} & { \tf   0.14} & { \tf  0.08} & { \tf  0.05} & { \tf      5.5}  & \\ 
{ \tf   17} & { \tf    HVC } & {\tf  001.7\mtm15.3+169} & { \tf 18 53.5} & { \tf \mtm34 26} & { \tf    169} & { \tf    175} & { \tf    126} & { \tf    85} & { \tf  0.3} & { \tf  0.2} & { \tf \mtm90} & { \tf   0.54} & { \tf  0.11} & { \tf  0.10} & { \tf     35.4}  & \\ 
{ \tf   18} & { \tf    HVC } & {\tf  001.7+05.1\mtm243} & { \tf 17 30.2} & { \tf \mtm24 47} & { \tf \mtm243} & { \tf \mtm236} & { \tf \mtm300} & { \tf    69} & { \tf  0.2} & { \tf  0.2} & { \tf \mtm40} & { \tf   0.20} & { \tf  0.08} & { \tf  0.09} & { \tf     11.5}  & \\ 
{ \tf   19} & { \tf    HVC } & {\tf  001.8+03.6\mtm209} & { \tf 17 36.0} & { \tf \mtm25 31} & { \tf \mtm209} & { \tf \mtm202} & { \tf \mtm265} & { \tf   112} & { \tf  0.5} & { \tf  0.3} & { \tf \mtm80} & { \tf   2.42} & { \tf  0.34} & { \tf  0.35} & { \tf    192.3}  & \\ 
{ \tf   20} & { \tf  HVC } & {\tf  001.9\mtm58.5\mtm170} & { \tf 22 30.8} & { \tf \mtm38 58} & { \tf \mtm170} & { \tf \mtm166} & { \tf \mtm168} & { \tf    40} & { \tf  0.2} & { \tf  0.1} & { \tf \mtm80} & { \tf   0.07} & { \tf  0.06} & { \tf  0.05} & { \tf      5.3}  & \\ 
{ \tf   21} & { \tf CHVC } & {\tf  002.2\mtm21.5\mtm182} & { \tf 19 22.8} & { \tf \mtm36 09} & { \tf \mtm182} & { \tf \mtm174} & { \tf \mtm218} & { \tf    38} & { \tf  0.3} & { \tf  0.2} & { \tf \mtm70} & { \tf   0.23} & { \tf  0.09} & { \tf  0.06} & { \tf     10.0}  & \\ 
{ \tf   22} & { \tf   :HVC } & {\tf  002.2+06.2\mtm214} & { \tf 17 27.6} & { \tf \mtm23 49} & { \tf \mtm214} & { \tf \mtm205} & { \tf \mtm269} & { \tf    37} & { \tf  0.2} & { \tf  0.2} & { \tf    70} & { \tf   0.27} & { \tf  0.15} & { \tf  0.11} & { \tf     12.9}  & \\ 
{ \tf   23} & { \tf  HVC } & {\tf  002.3\mtm44.5\mtm084} & { \tf 21 18.2} & { \tf \mtm39 59} & { \tf  \mtm84} & { \tf  \mtm78} & { \tf  \mtm96} & { \tf    35} & { \tf  0.3} & { \tf  0.2} & { \tf \mtm80} & { \tf   0.29} & { \tf  0.14} & { \tf  0.08} & { \tf     11.8}  & \\ 
{ \tf   24} & { \tf  HVC } & {\tf  002.3\mtm43.1\mtm118} & { \tf 21 11.0} & { \tf \mtm40 01} & { \tf \mtm118} & { \tf \mtm111} & { \tf \mtm132} & { \tf    31} & { \tf  0.4} & { \tf  0.3} & { \tf \mtm80} & { \tf   2.38} & { \tf  1.01} & { \tf  0.60} & { \tf    286.7}  & \\ 
{ \tf   25} & { \tf :HVC } & {\tf  002.3\mtm11.8\mtm196} & { \tf 18 38.9} & { \tf \mtm32 28} & { \tf \mtm196} & { \tf \mtm188} & { \tf \mtm240} & { \tf    36} & { \tf  0.7} & { \tf  0.3} & { \tf    80} & { \tf   1.67} & { \tf  0.21} & { \tf  0.17} & { \tf     89.9}  & \\ 
{ \tf   26} & { \tf  HVC } & {\tf  002.4\mtm07.5\mtm159} & { \tf 18 21.4} & { \tf \mtm30 30} & { \tf \mtm159} & { \tf \mtm150} & { \tf \mtm205} & { \tf    55} & { \tf  0.3} & { \tf  0.3} & { \tf    30} & { \tf   0.83} & { \tf  0.16} & { \tf  0.17} & { \tf     52.4}  & \\ 
{ \tf   27} & { \tf   CHVC } & {\tf  002.5+23.1\mtm138} & { \tf 16 30.4} & { \tf \mtm13 38} & { \tf \mtm138} & { \tf \mtm129} & { \tf \mtm198} & { \tf    31} & { \tf  0.2} & { \tf  0.1} & { \tf \mtm60} & { \tf   0.24} & { \tf  0.21} & { \tf  0.13} & { \tf     15.3}  & \\ 
{ \tf   28} & { \tf    HVC } & {\tf  002.6+07.1\mtm085} & { \tf 17 25.3} & { \tf \mtm22 56} & { \tf  \mtm85} & { \tf  \mtm75} & { \tf \mtm139} & { \tf    39} & { \tf  0.2} & { \tf  0.1} & { \tf \mtm60} & { \tf   0.42} & { \tf  0.31} & { \tf  0.23} & { \tf     24.7}  & \\ 
{ \tf   29} & { \tf  HVC } & {\tf  002.7\mtm69.5\mtm086} & { \tf 23 23.7} & { \tf \mtm35 44} & { \tf  \mtm86} & { \tf  \mtm82} & { \tf  \mtm70} & { \tf    37} & { \tf  0.3} & { \tf  0.2} & { \tf    80} & { \tf   0.70} & { \tf  0.38} & { \tf  0.28} & { \tf     63.7}  & \\ 
{ \tf   30} & { \tf    HVC } & {\tf  002.8+20.0\mtm118} & { \tf 16 40.9} & { \tf \mtm15 19} & { \tf \mtm118} & { \tf \mtm108} & { \tf \mtm176} & { \tf    30} & { \tf  0.2} & { \tf  0.2} & { \tf \mtm10} & { \tf   0.15} & { \tf  0.06} & { \tf  0.04} & { \tf      4.4}  & \\ 
\hline
\end{tabular}
\end{sideways}
\end{table}

\begin{table}
\begin{sideways}
\begin{tabular}{\tableallign}
\hline
\tableheader
\hline
{ \tf   31} & { \tf :HVC } & {\tf  002.9\mtm23.5\mtm120} & { \tf 19 33.2} & { \tf \mtm36 14} & { \tf \mtm120} & { \tf \mtm110} & { \tf \mtm151} & { \tf    37} & { \tf  0.2} & { \tf  0.2} & { \tf \mtm30} & { \tf   0.08} & { \tf  0.06} & { \tf  0.04} & { \tf      4.1}  & \\ 
{ \tf   32} & { \tf  HVC } & {\tf  002.9\mtm11.5\mtm145} & { \tf 18 39.0} & { \tf \mtm31 47} & { \tf \mtm145} & { \tf \mtm134} & { \tf \mtm185} & { \tf    37} & { \tf  0.8} & { \tf  0.7} & { \tf    20} & { \tf   0.68} & { \tf  0.08} & { \tf  0.07} & { \tf     19.5}  & \\ 
{ \tf   33} & { \tf  HVC } & {\tf  002.9\mtm06.2\mtm088} & { \tf 18 16.9} & { \tf \mtm29 29} & { \tf  \mtm88} & { \tf  \mtm77} & { \tf \mtm133} & { \tf    37} & { \tf  0.2} & { \tf  0.2} & { \tf \mtm50} & { \tf   0.31} & { \tf  0.31} & { \tf  0.22} & { \tf     22.3}  & \\ 
{ \tf   34} & { \tf  HVC } & {\tf  003.1\mtm59.8\mtm143} & { \tf 22 37.0} & { \tf \mtm38 04} & { \tf \mtm143} & { \tf \mtm137} & { \tf \mtm137} & { \tf    46} & { \tf  0.2} & { \tf  0.2} & { \tf    20} & { \tf   0.09} & { \tf  0.06} & { \tf  0.05} & { \tf      4.7}  & \\ 
{ \tf   35} & { \tf  HVC } & {\tf  003.2\mtm38.2\mtm123} & { \tf 20 46.2} & { \tf \mtm39 01} & { \tf \mtm123} & { \tf \mtm113} & { \tf \mtm138} & { \tf    38} & { \tf  0.2} & { \tf  0.2} & { \tf    60} & { \tf   0.43} & { \tf  0.28} & { \tf  0.20} & { \tf     31.1}  & \\ 
{ \tf   36} & { \tf CHVC } & {\tf  003.2\mtm26.7\mtm182} & { \tf 19 48.8} & { \tf \mtm36 51} & { \tf \mtm182} & { \tf \mtm171} & { \tf \mtm209} & { \tf    42} & { \tf  0.2} & { \tf  0.2} & { \tf \mtm50} & { \tf   0.13} & { \tf  0.07} & { \tf  0.05} & { \tf      5.3}  & \\ 
{ \tf   37} & { \tf :HVC } & {\tf  003.2\mtm14.2\mtm201} & { \tf 18 51.1} & { \tf \mtm32 43} & { \tf \mtm201} & { \tf \mtm190} & { \tf \mtm239} & { \tf    35} & { \tf  0.2} & { \tf  0.1} & { \tf     0} & { \tf   0.20} & { \tf  0.20} & { \tf  0.15} & { \tf     13.3}  & \\ 
{ \tf   38} & { \tf      HVC } & {\tf  003.3+10.1+106} & { \tf 17 16.1} & { \tf \mtm20 40} & { \tf    106} & { \tf    119} & { \tf     54} & { \tf    39} & { \tf  0.2} & { \tf  0.2} & { \tf \mtm50} & { \tf   1.02} & { \tf  0.15} & { \tf  0.09} & { \tf     25.5}  & \\ 
{ \tf   39} & { \tf    HVC } & {\tf  003.4+20.5\mtm118} & { \tf 16 40.7} & { \tf \mtm14 35} & { \tf \mtm118} & { \tf \mtm106} & { \tf \mtm174} & { \tf    39} & { \tf  0.2} & { \tf  0.2} & { \tf \mtm70} & { \tf   0.18} & { \tf  0.10} & { \tf  0.07} & { \tf      6.4}  & \\ 
{ \tf   40} & { \tf     :HVC } & {\tf  003.5+08.7+363} & { \tf 17 21.8} & { \tf \mtm21 19} & { \tf    363} & { \tf    377} & { \tf    313} & { \tf    34} & { \tf  0.2} & { \tf  0.2} & { \tf    10} & { \tf   0.09} & { \tf  0.09} & { \tf  0.05} & { \tf      3.4}  & \\ 
{ \tf   41} & { \tf CHVC } & {\tf  003.6\mtm65.6\mtm119} & { \tf 23 05.0} & { \tf \mtm36 31} & { \tf \mtm119} & { \tf \mtm113} & { \tf \mtm106} & { \tf    36} & { \tf  0.2} & { \tf  0.2} & { \tf \mtm60} & { \tf   0.26} & { \tf  0.17} & { \tf  0.12} & { \tf     13.7}  & \\ 
{ \tf   42} & { \tf  HVC } & {\tf  003.8\mtm43.0\mtm141} & { \tf 21 10.9} & { \tf \mtm38 54} & { \tf \mtm141} & { \tf \mtm131} & { \tf \mtm150} & { \tf    30} & { \tf  0.2} & { \tf  0.2} & { \tf    80} & { \tf   0.14} & { \tf  0.08} & { \tf  0.05} & { \tf      4.5}  & \\ 
{ \tf   43} & { \tf  HVC } & {\tf  004.1\mtm36.3\mtm130} & { \tf 20 36.8} & { \tf \mtm37 59} & { \tf \mtm130} & { \tf \mtm117} & { \tf \mtm144} & { \tf    35} & { \tf  0.4} & { \tf  0.2} & { \tf    40} & { \tf   2.28} & { \tf  0.62} & { \tf  0.42} & { \tf    259.6}  & \\ 
{ \tf   44} & { \tf    HVC } & {\tf  004.1+22.5\mtm139} & { \tf 16 35.8} & { \tf \mtm12 49} & { \tf \mtm139} & { \tf \mtm124} & { \tf \mtm192} & { \tf    52} & { \tf  0.2} & { \tf  0.2} & { \tf \mtm10} & { \tf   0.13} & { \tf  0.07} & { \tf  0.05} & { \tf      5.4}  & \\ 
{ \tf   45} & { \tf      HVC } & {\tf  004.2+05.5+211} & { \tf 17 34.7} & { \tf \mtm22 29} & { \tf    211} & { \tf    227} & { \tf    165} & { \tf    44} & { \tf  0.4} & { \tf  0.1} & { \tf    40} & { \tf   0.81} & { \tf  0.21} & { \tf  0.18} & { \tf     49.9}  & \\ 
{ \tf   46} & { \tf  HVC } & {\tf  004.4\mtm25.6\mtm086} & { \tf 19 44.8} & { \tf \mtm35 30} & { \tf  \mtm86} & { \tf  \mtm71} & { \tf \mtm109} & { \tf    24} & { \tf  0.2} & { \tf  0.2} & { \tf    70} & { \tf   0.11} & { \tf  0.07} & { \tf  0.03} & { \tf      4.3}  & \\ 
{ \tf   47} & { \tf   GLXY } & {\tf  004.5\mtm77.2+209} & { \tf 23 57.9} & { \tf \mtm32 35} & { \tf    209} & { \tf    213} & { \tf    234} & { \tf   165} & { \tf  0.2} & { \tf  0.1} & { \tf \mtm10} & { \tf   0.54} & { \tf  1.11} & { \tf  2.93} & { \tf    310.9}  & \\ 
{ \tf   48} & { \tf :HVC } & {\tf  004.5\mtm63.8\mtm116} & { \tf 22 55.9} & { \tf \mtm36 35} & { \tf \mtm116} & { \tf \mtm109} & { \tf \mtm103} & { \tf    46} & { \tf  0.2} & { \tf  0.2} & { \tf    40} & { \tf   0.43} & { \tf  0.14} & { \tf  0.12} & { \tf     25.5}  & \\ 
{ \tf   49} & { \tf  HVC } & {\tf  004.6\mtm61.0\mtm101} & { \tf 22 42.0} & { \tf \mtm37 07} & { \tf \mtm101} & { \tf  \mtm92} & { \tf  \mtm90} & { \tf    38} & { \tf  1.3} & { \tf  0.5} & { \tf    10} & { \tf   1.05} & { \tf  0.18} & { \tf  0.15} & { \tf     69.5}  & \\ 
{ \tf   50} & { \tf  HVC } & {\tf  004.6\mtm09.7\mtm086} & { \tf 18 34.6} & { \tf \mtm29 32} & { \tf  \mtm86} & { \tf  \mtm68} & { \tf \mtm120} & { \tf    37} & { \tf  0.2} & { \tf  0.2} & { \tf    40} & { \tf   0.10} & { \tf  0.08} & { \tf  0.05} & { \tf      4.3}  & \\ 
{ \tf   51} & { \tf    HVC } & {\tf  004.8\mtm06.2+206} & { \tf 18 21.0} & { \tf \mtm27 52} & { \tf    206} & { \tf    224} & { \tf    170} & { \tf    48} & { \tf  0.2} & { \tf  0.1} & { \tf    30} & { \tf   0.57} & { \tf  0.13} & { \tf  0.12} & { \tf     57.2}  & \\ 
{ \tf   52} & { \tf  HVC } & {\tf  005.0\mtm66.7\mtm123} & { \tf 23 09.7} & { \tf \mtm35 42} & { \tf \mtm123} & { \tf \mtm115} & { \tf \mtm106} & { \tf    59} & { \tf  0.3} & { \tf  0.1} & { \tf \mtm50} & { \tf   0.20} & { \tf  0.05} & { \tf  0.05} & { \tf      9.7}  & \\ 
{ \tf   53} & { \tf    HVC } & {\tf  005.0+04.9\mtm156} & { \tf 17 38.5} & { \tf \mtm22 07} & { \tf \mtm156} & { \tf \mtm137} & { \tf \mtm198} & { \tf    19} & { \tf  0.7} & { \tf  0.5} & { \tf \mtm50} & { \tf   0.50} & { \tf  0.14} & { \tf  0.05} & { \tf     12.4}  & \\ 
{ \tf   54} & { \tf    HVC } & {\tf  005.3\mtm63.6+091} & { \tf 22 54.3} & { \tf \mtm36 18} & { \tf     91} & { \tf    100} & { \tf    105} & { \tf    39} & { \tf  0.5} & { \tf  0.2} & { \tf \mtm30} & { \tf   1.06} & { \tf  0.19} & { \tf  0.14} & { \tf     57.0}  & \\ 
{ \tf   55} & { \tf  HVC } & {\tf  005.3\mtm35.9\mtm164} & { \tf 20 35.6} & { \tf \mtm37 01} & { \tf \mtm164} & { \tf \mtm148} & { \tf \mtm174} & { \tf    35} & { \tf  0.2} & { \tf  0.2} & { \tf \mtm50} & { \tf   0.13} & { \tf  0.08} & { \tf  0.05} & { \tf      4.8}  & \\ 
{ \tf   56} & { \tf  HVC } & {\tf  005.3\mtm34.1\mtm101} & { \tf 20 26.6} & { \tf \mtm36 41} & { \tf \mtm101} & { \tf  \mtm84} & { \tf \mtm112} & { \tf    38} & { \tf  0.2} & { \tf  0.2} & { \tf \mtm20} & { \tf   0.51} & { \tf  0.34} & { \tf  0.27} & { \tf     38.7}  & \\ 
{ \tf   57} & { \tf  HVC } & {\tf  005.3\mtm21.4\mtm130} & { \tf 19 26.6} & { \tf \mtm33 23} & { \tf \mtm130} & { \tf \mtm111} & { \tf \mtm152} & { \tf    43} & { \tf  0.2} & { \tf  0.2} & { \tf \mtm70} & { \tf   0.31} & { \tf  0.12} & { \tf  0.09} & { \tf     16.8}  & \\ 
{ \tf   58} & { \tf  HVC } & {\tf  005.4\mtm43.5\mtm115} & { \tf 21 13.8} & { \tf \mtm37 44} & { \tf \mtm115} & { \tf \mtm100} & { \tf \mtm118} & { \tf    41} & { \tf  0.5} & { \tf  0.2} & { \tf \mtm40} & { \tf   0.68} & { \tf  0.09} & { \tf  0.07} & { \tf     24.6}  & \\ 
{ \tf   59} & { \tf    HVC } & {\tf  005.4+27.2\mtm119} & { \tf 16 23.5} & { \tf \mtm09 02} & { \tf \mtm119} & { \tf \mtm101} & { \tf \mtm168} & { \tf    42} & { \tf  0.3} & { \tf  0.2} & { \tf \mtm40} & { \tf   0.20} & { \tf  0.06} & { \tf  0.05} & { \tf      6.7}  & \\ 
{ \tf   60} & { \tf  HVC } & {\tf  005.5\mtm50.7\mtm128} & { \tf 21 50.2} & { \tf \mtm37 45} & { \tf \mtm128} & { \tf \mtm114} & { \tf \mtm124} & { \tf    35} & { \tf  0.3} & { \tf  0.2} & { \tf \mtm20} & { \tf   0.66} & { \tf  0.18} & { \tf  0.12} & { \tf     38.6}  & \\ 
\hline
\end{tabular}
\end{sideways}
\end{table}

\clearpage

\renewcommand{\arraystretch}{0.90}

\begin{table}
\caption[]{Median properties of cataloged objects (Dec. $< 2$\deg, $-500 < V_{\rm LSR} < +500$ \kms)}
\label{tab:hvc}
\begin{tabular}{rrrrrrrrrrrrrl}
\tableline
\tableline
\multicolumn{1}{c}{                                             Class} &
\multicolumn{1}{c}{                                               \#} &
\multicolumn{1}{c}{     $\scriptstyle  V_{\rm LSR}$} &
\multicolumn{1}{c}{     $\scriptstyle  V_{\rm GSR}$} &
\multicolumn{1}{c}{    $\scriptstyle  V_{\rm LGSR}$} &
\multicolumn{1}{c}{                                          \tf FWHM} &
\multicolumn{1}{c}{      $\scriptstyle N_{\scriptscriptstyle \rm HI}$} &
\multicolumn{1}{c}{    $\scriptstyle T_{\scriptscriptstyle \rm peak}$} &
\multicolumn{1}{c}{                           $\scriptstyle \rm Flux$} &
\multicolumn{1}{c}{                           $\scriptstyle \rm Size$} &
\multicolumn{1}{c}{                                           \tf MAJ} &
\multicolumn{1}{c}{                                           \tf MIN}  \\
\multicolumn{1}{c}{                                                \ } &
\multicolumn{1}{c}{                                                \ } &
\multicolumn{1}{c}{                    $\scriptstyle (\rm km~s^{-1})$} &
\multicolumn{1}{c}{                    $\scriptstyle (\rm km~s^{-1})$} &
\multicolumn{1}{c}{                    $\scriptstyle (\rm km~s^{-1})$} &
\multicolumn{1}{c}{                    $\scriptstyle (\rm km~s^{-1})$} &
\multicolumn{1}{c}{              $\scriptstyle (10^{20} \rm cm^{-2})$} &
\multicolumn{1}{c}{                            $\scriptstyle (\rm K)$} &
\multicolumn{1}{c}{               $\scriptstyle (\rm Jy\; km~s^{-1})$} &
\multicolumn{1}{c}{                      $\scriptstyle (\rm deg^{2})$} &
\multicolumn{1}{c}{                          $\scriptstyle (\rm deg)$} &
\multicolumn{1}{c}{                          $\scriptstyle (\rm deg)$} \\

\tableline
HVCs & 1618 & 117 & -30 & -62 & 36 & 0.08 & 0.12 & 19.0 & 0.42 & 0.3 & 0.2  \\
CHVCs & 179 &  95 & -38 & -57 & 35 & 0.14 & 0.2 & 19.9 & 0.36 & 0.2 & 0.2 \\
:HVCs & 159 & 105 & -22 & -66 & 36 & 0.12 & 0.16 & 17.0 & 0.35 & 0.2 & 0.2 \\
All Clouds & 1956 & 114 & -30 & -63 & 36  & 0.09 & 0.12 & 19.0 & 0.41 & 0.3 & 0.2\\
GLXYs & 41  & 386 & 264 & 218 & 57 & 0.33 & 0.34 & 55.5 & 0.28 & 0.2 & 0.1\\
\tableline
\\
\end{tabular}

\end{table}

\begin{table}
\renewcommand{\arraystretch}{0.6}
\caption[]{The beginning of the catalog of XHVCs, objects which have high-velocity emission
($-60$ \kms $> V_{\rm dev} > 60$ \kms), but merge with Galactic emission
below this velocity range.  The complete XHVC table
includes 4 galaxies and is available at {\it
http://wwwatnf.atnf.csiro.au/research/multibeam} and the electronic version of this journal.}
\begin{sideways}
\begin{tabular}{\tableallign}
\hline
\tableheader
\hline
{\tf  1} & {\tf   XHVC } & {\tf 000.2+07.0\mtm095} & {\tf17 19.6} & {\tf\mtm25 01} & {\tf \mtm95} & {\tf \mtm94} & {\tf\mtm160} & {\tf 39} & {\tf 0.6} & {\tf 0.4} & {\tf   83} & {\tf  6.49} & {\tf  1.73} & {\tf 1.23} & {\tf          1506.2}  & \\ 
{\tf  2} & {\tf   XHVC } & {\tf 001.3+05.0\mtm089} & {\tf17 29.8} & {\tf\mtm25 13} & {\tf \mtm89} & {\tf \mtm85} & {\tf\mtm148} & {\tf 34} & {\tf 0.5} & {\tf 0.3} & {\tf   56} & {\tf  1.68} & {\tf  1.19} & {\tf 0.73} & {\tf           365.0}  & \\ 
{\tf  3} & {\tf   XHVC } & {\tf 002.7\mtm03.1+132} & {\tf18 04.0} & {\tf\mtm28 11} & {\tf   132} & {\tf   143} & {\tf    85} & {\tf 40} & {\tf 0.2} & {\tf 0.2} & {\tf\mtm37} & {\tf  0.81} & {\tf  0.72} & {\tf 0.57} & {\tf           122.1}  & \\ 
{\tf  4} & {\tf       XHVC } & {\tf 004+05\mtm095} & {\tf17 35.2} & {\tf\mtm23 17} & {\tf \mtm95} & {\tf \mtm81} & {\tf\mtm143} & {\tf 29} & {\tf12.1} & {\tf 8.3} & {\tf\mtm48} & {\tf  1.28} & {\tf  0.68} & {\tf 0.33} & {\tf           124.3}  & \\ 
{\tf  5} & {\tf   XHVC } & {\tf 004.3+05.1\mtm097} & {\tf17 36.6} & {\tf\mtm22 34} & {\tf \mtm97} & {\tf \mtm81} & {\tf\mtm142} & {\tf 36} & {\tf 0.6} & {\tf 0.3} & {\tf\mtm44} & {\tf  2.66} & {\tf  1.14} & {\tf 0.98} & {\tf           555.7}  & \\ 
{\tf  6} & {\tf   XHVC } & {\tf 005.9+08.9\mtm114} & {\tf17 26.3} & {\tf\mtm19 10} & {\tf\mtm114} & {\tf \mtm91} & {\tf\mtm154} & {\tf 35} & {\tf 0.4} & {\tf 0.3} & {\tf\mtm41} & {\tf  1.59} & {\tf  0.16} & {\tf 0.11} & {\tf            58.8}  & \\ 
{\tf  7} & {\tf   XHVC } & {\tf 010.6\mtm14.2+092} & {\tf19 04.2} & {\tf\mtm26 06} & {\tf    92} & {\tf   131} & {\tf    88} & {\tf 52} & {\tf 0.2} & {\tf 0.2} & {\tf   73} & {\tf  0.24} & {\tf  0.48} & {\tf 0.28} & {\tf            20.1}  & \\ 
{\tf  8} & {\tf   XHVC } & {\tf 012.4+15.6\mtm094} & {\tf17 17.3} & {\tf\mtm10 15} & {\tf \mtm94} & {\tf \mtm48} & {\tf\mtm107} & {\tf 32} & {\tf 0.2} & {\tf 0.2} & {\tf   66} & {\tf  0.65} & {\tf  0.78} & {\tf 0.49} & {\tf            85.4}  & \\ 
{\tf  9} & {\tf   XHVC } & {\tf 013.4+07.0\mtm101} & {\tf17 49.2} & {\tf\mtm13 53} & {\tf\mtm101} & {\tf \mtm50} & {\tf\mtm105} & {\tf 33} & {\tf 0.4} & {\tf 0.3} & {\tf\mtm16} & {\tf  1.12} & {\tf  0.96} & {\tf 0.57} & {\tf           196.0}  & \\ 
{\tf 10} & {\tf XHVC } & {\tf 015.1\mtm58.2\mtm116} & {\tf22 26.4} & {\tf\mtm32 05} & {\tf\mtm116} & {\tf \mtm86} & {\tf \mtm83} & {\tf 87} & {\tf 0.2} & {\tf 0.2} & {\tf\mtm41} & {\tf  0.36} & {\tf  0.18} & {\tf 0.22} & {\tf            29.8}  & \\ 
{\tf 11} & {\tf   XHVC } & {\tf  017.8+07.2\mtm079} & {\tf17 57.6} & {\tf\mtm10 01} & {\tf \mtm79} & {\tf \mtm12} & {\tf \mtm63} & {\tf 35} & {\tf 0.5} & {\tf 0.2} & {\tf \mtm7} & {\tf  4.61} & {\tf  1.25} & {\tf 0.92} & {\tf          1049.4}  & \\ 
{\tf 12} & {\tf   XHVC } & {\tf  017.8+15.5\mtm085} & {\tf17 28.3} & {\tf\mtm05 50} & {\tf \mtm85} & {\tf \mtm20} & {\tf \mtm74} & {\tf 59} & {\tf 0.2} & {\tf 0.2} & {\tf\mtm61} & {\tf  5.13} & {\tf  0.97} & {\tf 0.64} & {\tf           683.6}  & \\ 
{\tf 13} & {\tf       XHVC } & {\tf  018\mtm11+130} & {\tf19 04.2} & {\tf\mtm18 38} & {\tf   130} & {\tf   195} & {\tf   156} & {\tf 39} & {\tf11.2} & {\tf 5.7} & {\tf   17} & {\tf  0.83} & {\tf  0.27} & {\tf 0.19} & {\tf            53.0}  & \\ 
{\tf 14} & {\tf   XHVC } & {\tf  018.7+12.8\mtm095} & {\tf17 39.4} & {\tf\mtm06 29} & {\tf \mtm95} & {\tf \mtm26} & {\tf \mtm79} & {\tf 33} & {\tf 1.6} & {\tf 1.0} & {\tf \mtm5} & {\tf 26.86} & {\tf  2.05} & {\tf 1.45} & {\tf          7301.6}  & \\ 
{\tf 15} & {\tf     XHVC } & {\tf  019.4+15.0+106} & {\tf17 33.2} & {\tf\mtm04 49} & {\tf   106} & {\tf   177} & {\tf   124} & {\tf 35} & {\tf 0.2} & {\tf 0.1} & {\tf   80} & {\tf  0.21} & {\tf  0.38} & {\tf 0.24} & {\tf            16.0}  & \\ 
{\tf 16} & {\tf   XHVC } & {\tf  022.6\mtm22.6+089} & {\tf19 57.0} & {\tf\mtm18 48} & {\tf    89} & {\tf   167} & {\tf   142} & {\tf 31} & {\tf 0.5} & {\tf 0.2} & {\tf   54} & {\tf  3.19} & {\tf  0.61} & {\tf 0.40} & {\tf           297.9}  & \\ 
{\tf 17} & {\tf XHVC } & {\tf  025.6\mtm46.8\mtm108} & {\tf21 38.9} & {\tf\mtm24 13} & {\tf\mtm108} & {\tf \mtm43} & {\tf \mtm44} & {\tf 34} & {\tf 0.5} & {\tf 0.4} & {\tf    6} & {\tf  1.70} & {\tf  0.44} & {\tf 0.27} & {\tf           166.5}  & \\ 
{\tf 18} & {\tf     XHVC } & {\tf  027.0+10.9+115} & {\tf18 01.7} & {\tf\mtm00 15} & {\tf   115} & {\tf   213} & {\tf   170} & {\tf 41} & {\tf 0.5} & {\tf 0.2} & {\tf   55} & {\tf  1.82} & {\tf  0.72} & {\tf 0.56} & {\tf           272.3}  & \\ 
{\tf 19} & {\tf XHVC } & {\tf  027.5\mtm48.6\mtm105} & {\tf21 48.2} & {\tf\mtm23 26} & {\tf\mtm105} & {\tf \mtm38} & {\tf \mtm36} & {\tf 35} & {\tf 0.5} & {\tf 0.4} & {\tf   75} & {\tf  2.49} & {\tf  0.56} & {\tf 0.37} & {\tf           323.2}  & \\ 
{\tf 20} & {\tf   XHVC } & {\tf  027.5\mtm23.5+135} & {\tf20 07.5} & {\tf\mtm15 02} & {\tf   135} & {\tf   228} & {\tf   209} & {\tf 37} & {\tf 0.7} & {\tf 0.3} & {\tf   82} & {\tf  1.28} & {\tf  0.26} & {\tf 0.19} & {\tf           113.5}  & \\ 
{\tf 21} & {\tf   XHVC } & {\tf  028.1\mtm16.4+100} & {\tf19 42.1} & {\tf\mtm11 35} & {\tf   100} & {\tf   199} & {\tf   175} & {\tf 35} & {\tf 0.3} & {\tf 0.2} & {\tf\mtm27} & {\tf  0.92} & {\tf  0.52} & {\tf 0.33} & {\tf            91.7}  & \\ 
{\tf 22} & {\tf   XHVC } & {\tf  031.1\mtm12.1+105} & {\tf19 31.2} & {\tf\mtm07 07} & {\tf   105} & {\tf   216} & {\tf   191} & {\tf 37} & {\tf 0.9} & {\tf 0.3} & {\tf   57} & {\tf  3.64} & {\tf  0.93} & {\tf 0.79} & {\tf           615.3}  & \\ 
{\tf 23} & {\tf   XHVC } & {\tf  035.8\mtm10.0+129} & {\tf19 32.0} & {\tf\mtm02 02} & {\tf   129} & {\tf   255} & {\tf   235} & {\tf 55} & {\tf 0.8} & {\tf 0.2} & {\tf   63} & {\tf  1.74} & {\tf  1.31} & {\tf 0.99} & {\tf           350.2}  & \\ 
{\tf 24} & {\tf   XHVC } & {\tf  039.9\mtm21.2+090} & {\tf20 19.3} & {\tf\mtm03 41} & {\tf    90} & {\tf   222} & {\tf   214} & {\tf 36} & {\tf 1.3} & {\tf 0.5} & {\tf\mtm82} & {\tf  7.24} & {\tf  2.40} & {\tf 1.94} & {\tf          2274.5}  & \\ 
{\tf 25} & {\tf XHVC } & {\tf  045.5\mtm81.9\mtm105} & {\tf00 16.4} & {\tf\mtm25 05} & {\tf\mtm105} & {\tf \mtm83} & {\tf \mtm51} & {\tf 39} & {\tf 0.5} & {\tf 0.4} & {\tf   48} & {\tf 57.33} & {\tf  4.93} & {\tf 4.53} & {\tf         28086.3}  & \\ 
{\tf 26} & {\tf   XHVC } & {\tf  047.0\mtm28.7+103} & {\tf20 58.2} & {\tf\mtm01 38} & {\tf   103} & {\tf   244} & {\tf   249} & {\tf 37} & {\tf 0.5} & {\tf 0.3} & {\tf\mtm42} & {\tf  0.80} & {\tf  0.37} & {\tf 0.24} & {\tf            88.8}  & \\ 
{\tf 27} & {\tf   XHVC } & {\tf  048.6\mtm30.1+104} & {\tf21 05.9} & {\tf\mtm01 11} & {\tf   104} & {\tf   247} & {\tf   255} & {\tf 36} & {\tf 0.4} & {\tf 0.2} & {\tf\mtm57} & {\tf  0.68} & {\tf  0.59} & {\tf 0.42} & {\tf           114.9}  & \\ 
{\tf 28} & {\tf XHVC } & {\tf  061.6\mtm62.0\mtm079} & {\tf23 12.0} & {\tf\mtm11 39} & {\tf \mtm79} & {\tf    12} & {\tf    46} & {\tf 29} & {\tf 0.7} & {\tf 0.7} & {\tf \mtm3} & {\tf  4.25} & {\tf  0.61} & {\tf 0.38} & {\tf           462.0}  & \\ 
{\tf 29} & {\tf GLXY } & {\tf  095.1\mtm88.0+344} & {\tf00 47.2} & {\tf\mtm25 20} & {\tf   344} & {\tf   352} & {\tf   388} & {\tf105} & {\tf 0.2} & {\tf 0.1} & {\tf\mtm85} & {\tf  0.58} & {\tf  2.23} & {\tf 5.09} & {\tf           479.6}  & {\tf  NGC\ 253}  \\ 
{\tf 30} & {\tf XHVC } & {\tf  197.8\mtm24.4\mtm084} & {\tf04 57.8} & {\tf  +01 22} & {\tf \mtm84} & {\tf\mtm146} & {\tf \mtm88} & {\tf 46} & {\tf 0.3} & {\tf 0.3} & {\tf\mtm80} & {\tf  0.65} & {\tf  0.68} & {\tf 0.44} & {\tf           110.8}  & \\ 
\hline
\end{tabular}
\end{sideways}
\end{table}

\label{tab:catalog}


\begin{thebibliography}{} 
\bibitem[1987]{baja87}
Bajaja, E., Morras, R., \& P\"oppel, W.G.L. 1987, Pub. Astr. Inst. Czech. Ac. Sci., 69, 237 
\bibitem[2001]{barn00}
Barnes, D.G., Staveley-Smith, L., de Blok, W.J.G., et al. 2001, MNRAS, 322, 486
\bibitem[1999]{blit99}
Blitz, L., Spergel, D.N., Teuben, P.J., Hartmann, D., \& Burton, W.B. 1999, ApJ, 514, 818 
\bibitem[1999]{brau99}
Braun, R., \& Burton, W.B. 1999, A\&A, 341, 437 (BB99)
\bibitem[2000]{brau00}
Braun R., \& Burton W.B. 2000, A\&A, 354, 853 
\bibitem[2000]{brun00} 
Br\"uns, C., Kerp, J., Kalberla, P.M.W., \& Mebold, U. 2000, A\&A, 357, 120
\bibitem[1988]{burt98}
Burton, W.B. 1988, in Galactic \& Extragalactic Radio Astronomy,
eds. Verschuur, G.L. \& Kellermann, K.I., Springer-Verlag, 295
\bibitem[2001]{burt01}
Burton, W.B., Braun, R., \& Chengalur, J.N. 2001, A\&A, submitted (astro-ph/0102085)
\bibitem[1999]{dave99}
Dave, R., Hernquist, L., Katz, N., \& Weinberg, D.H. 1999, ApJ, 511, 521
\bibitem[2001]{dehe00}
de Heij, V., Braun, R., \& Burton, W.B. 2001, A\&A, in preparation
\bibitem[1976]{eich76}
Eichler, D. 1976, ApJ, 208, 694
\bibitem[1976]{eina76}
Einasto, J., Haud, U., J\^oeveer, M., \& Kaasik, A. 1976, MNRAS, 177, 357
\bibitem[2000]{gibs00}
Gibson, B.K., Giroux, M.L., Penton, S.V., Putman, M.E., Stocke, J.T.,
Shull, J.M. 2000, AJ, 120, 1830
\bibitem[1981]{giov81}
Giovanelli, R. 1981, AJ, 86, 1468 
\bibitem[2000]{gned00}
Gnedin, N.Y. 2000, ApJ, 542, 535
\bibitem[1993]{hami93}
Hamilton, A.J.S. 1993, ApJ, 417, 19
\bibitem[1997]{hart97}
Hartmann, D., \& Burton, W.B. 1997, ``Atlas of Galactic Neutral Hydrogen'' 
(Cambridge: Cambridge University Press)
\bibitem[2000]{henn00}
Henning, P.A. 2000, in ASP Conf. Ser. 219, Mapping the Hidden Universe, ed.
R.C. Kraan--Korteweg, P.A. Henning, \& H. Andernach
\bibitem[1998]{henn98}
Henning, P.A., Kraan--Korteweg, R.C., Rivers, A.J., Loan, A.J., Lahav, O.,
\& Burton, W.B. 1998, AJ, 115, 584
\bibitem[2001]{henn01}
Henning, P.A., Staveley--Smith, L., Ekers, R.D., Green, A.J., Haynes, R.F.,
Juraszek, S., Kesteven, M.J., Koribalski, B., Kraan--Korteweg, R.C., Price,
R.M., Sadler, E.M., \& Schr\"oder, A. 2001, AJ, in press (astro--ph/0003245)

\bibitem[1995]{hu1995}
Hu E. M., Kim T.-S., Cowie L. L., Songaila A., \& Rauch M. 1995, AJ, 110, 1526
\bibitem[1968]{huls68}
Hulsbosch, A.N.M. 1968, Bull. Astr. Inst. Netherlands, 20, 33

\bibitem[2000]{kilb00}
Kilborn, V. 2000, PhD Thesis, University of Melbourne
\bibitem[1999]{kilb99}
Kilborn, V., Webster, R. L., \& Staveley--Smith, L. 1999, PASA, 16, 8
\bibitem[1999]{klyp99}
Klypin, A., Kravtsov, A.V., Valenzuela, O., \& Prada, F. 1999, ApJ, 522, 82
\bibitem[2001{kori01}
Koribalski, B., et al.  2001, AJ, in preparation
\bibitem[1976]{mira76}
Mirabel, I.F., \& Franco, M.L. 1976, Ap\&SS, 39, 415

\bibitem[1999]{moor99}
Moore, B., Ghigna, S., Governato, G., Lake, G., Quinn, T., Stadel, J., \& Tozzi, P.
1999, ApJ, 524, L19
\bibitem[2000]{morr00}
Morras, R., Bajaja, E., Arnal, E.M., \& P\"oppel, W.G.L. 2000, A\&AS, 142, 25
\bibitem[2000]{mulc00}
Mulchaey, J.S. 2000, ARA\&A, 38, 289
\bibitem[1995]{murp95}
Murphy, E.M., Lockman, F., \& Savage, B.D. 1995, ApJ, 447, 642 
\bibitem[1966]{oort66}
Oort, J.H. 1966, Bull. Astr. Inst. Netherlands, 18, 421 
\bibitem[1968]{oort68}
Oort, J.H. 1968, in Proc. IAU Symp., 29, 41
\bibitem[1970]{oort70}
Oort, J.H. 1970, A\&A, 7, 381 
\bibitem[1981]{oort81}
Oort, J.H. 1981, A\&A, 94, 359 
\bibitem[2000]{pent00}
Penton, S.V., Shull, J.M., \& Stocke, J.T. 2000, ApJ, 544, 150
\bibitem[2001]{putm01}
Putman, M.E., et al. 2002, AJ, in prep
\bibitem[2000]{putm00}
Putman, M.E. 2000, PhD Thesis, Australian National University
\bibitem[1999]{putm99}
Putman, M.E., \& Gibson B.K. 1999, PASA, 16, 70
\bibitem[1998]{putm98}
Putman, M.E., et al. 1998, Nature, 394, 752 
\bibitem[2001]{quil01}
Quilis, V. \& Moore, B. 2001, ApJ, 555, L95
\bibitem[1986]{rees86}
Rees, M.J. 1986, MNRAS, 218, 25
\bibitem[2000]{rive00}
Rivers, A.J. 2000, PhD Thesis, University of New Mexico, Albuquerque
\bibitem[2001]{sanc01}
Sancisi, R., Fraternali, F., Oosterloo, T. \& van Moorsel, G. 2001, in 
Gas and Galaxy Evolution, ASP Conf. Series, in press (astro-ph/0009119)
\bibitem[2000]{schn00}
Schneider S.E., \& Rosenberg J.L. 2000, astro-ph/0010375
\bibitem[1974]{sara74}
Saraber, M.J.M., \& Shane, W.W. 1974, A\&A, 30, 365
\bibitem[2000]{semb00}
Sembach, K.R., et al. 2000, ApJ, 538, 31
\bibitem[1998]{stop98}
Stoppelenburg, P.S., Schwarz, U.J., \& van Woerden, H. 1998, A\&A, 338, 200
\bibitem[1993]{tayl93}
Taylor, J.H. \& Cordes, J.M. 1993, ApJ, 411, 674
\bibitem[1998]{tuft98}
Tufte, S.L., Reynolds, R.J., \& Haffner, L.M. 1998, ApJ, 504, 773
\bibitem[1999]{woer99}
van Woerden, H., Schwarz, U.J., Peletier, R.F., Wakker, B.P., \& Kalberla, P.M.W. 1999, Nature, 400, 138
\bibitem[1975]{vers75}
Verschuur, G.L. 1975, ARA\&A, 13, 257
\bibitem[1991]{wakk91a}
Wakker, B.P. 1991, A\&A, 250, 499
\bibitem[1999]{wakk99}
Wakker, B.P., Howk, J.C., Savage, B.D., et al. 1999, Nature, 402, 388
\bibitem[2001]{wakk01a}
Wakker, B.P., Kalberla, P.M.W., van Woerden, H., de Boer, K.S., \& Putman, M.E. 2001, AJ, in press
\bibitem[2001]{wakk01b}
Wakker, B.P., Oosterloo, T., \& Putman, M.E. 2001, ApJ, accepted
\bibitem[1991]{wakk91b}
Wakker, B.P., \& Schwarz, U. 1991, A\&A, 250, 484
\bibitem[1991]{wakk91c}
Wakker, B.P., \& van Woerden, H. 1991, A\&A, 250, 509
\bibitem[1997]{wakk97}
Wakker, B.P., \& van Woerden, H. 1997, ARA\&A, 35, 217
\bibitem[1972]{wawx72}
Wannier, P., \& Wrixon, G.T. 1972, ApJ, 173, L119
\bibitem[1972]{wann72}
Wannier, P., Wrixon, G.T., \& Wilson, R.W. 1972, A\&A, 18, 224
\bibitem[1999]{zwaa99}
Zwaan, M.A., Verheijen, M.A.W., \& Briggs, F.H. 1999, PASA, 16, 100
\end{thebibliography}
\end{document}